# Ultrafast Triplet Pair Formation and Subsequent Thermally Activated Dissociation Control Efficient Endothermic Singlet Exciton Fission

Hannah L. Stern*[1], Alexandre Cheminal[1], Shane R. Yost[2,3], Katharina Broch[1], Sam L. Bayliss[1], Kai Chen[4,5], Maxim Tabachnyk[1], Karl Thorley[6], Neil Greenham[1], Justin Hodgkiss[4,5], John Anthony[6], Martin Head-Gordon[2,3], Andrew J. Musser[1], Akshay Rao[1] and Richard H. Friend[1].

[1] Cavendish Laboratory, University of Cambridge, UK.

[2] Kenneth S. Pitzer Center for Theoretical Chemistry, Department of Chemistry, University of California, Berkeley, USA.

[3] Chemical Science Division, Lawrence Berkeley National Laboratory, Berkeley, USA.

[4] MacDiarmid Institute for Advanced Materials and Nanotechnology, New Zealand.

[5] School of Chemical and Physical Sciences, Victoria University of Wellington, New Zealand.

[6] University of Kentucky, Lexington, USA.

**Abstract**

Singlet exciton fission (SF), the conversion of one spin-singlet exciton ($S_1$) into two spin-triplet excitons ($T_1$), could provide a means to overcome the Shockley-Queisser limit in photovoltaics. SF as measured by the decay of $S_1$ has been shown to occur efficiently and independently of temperature even when the energy of $S_1$ is as much as 200 meV less than $2T_1$. Here, we study films of TIPS-tetracene using transient optical spectroscopy and show that the initial rise of the triplet pair state (TT) occurs in 300 fs, matched by rapid loss of $S_1$ stimulated emission, and that this process is mediated by the strong coupling of electronic and vibrational degrees of freedom. This is followed by a slower 10 ps morphology-dependent phase of $S_1$ decay and TT growth. We observe the TT to be thermally dissociated on 10-100 ns timescales to form free triplets. This provides a model for 'temperature independent', efficient TT formation and thermally activated TT separation.

Singlet exciton fission (SF) is a quantum mechanical phenomenon unique to organic chromophores that could provide a route to breaking the Shockley-Queisser limit on the efficiency of single junction photovoltaics (PVs)[1,2,3]. In this process, a photogenerated spin-0 singlet exciton ($S_1$) is converted to two spin-1 triplet excitons ($T_1$). It has been proposed that this conversion is mediated by a triplet pair intermediate state (TT), which forms an overall spin-0 state[4]. This means that SF does not require a spin flip and can proceed on <100 fs timescales when SF is exothermic, i.e $E(S_1)>2E(T_1)$, allowing for near unity efficiency (200% triplet yield) in materials such as pentacene[5].

But intriguingly, SF also proceeds very efficiently in endothermic systems, where $E(S_1)<2E(T_1)$, overcoming energy barriers ($E_b = 2E(T_1) - E(S_1)$) of up to 200 meV[6]. Such systems are of particular technological importance, as most of the materials with $E(T_1)$ comparable to the bandgap of silicon (1.1eV) fall into this category, including perylenediimides[7] and acenes such as tetracene ($E_b \cong 180$ meV), which is the most well studied endothermic SF system[6,8,9]. Extensive work by Bardeen and others has unambiguously shown that free triplets are produced in a high yield in polycrystalline tetracene, yet $S_1$ decays independently of temperature on a 70-90 ps timescale [6,10,11,12]. The rate of decay is three orders of magnitude slower than in pentacene, despite similar electronic couplings between the relevant electronic states[13]. So, what controls the decay rate of $S_1$ and how can this state efficiently overcome an endothermic barrier to generate free triplets? The answer to this question is both of fundamental interest and of practical importance since it could allow for greatly enhanced efficiency in photovoltaic devices, which effectively harvest energy from the environment via endothermic SF.

To investigate the mechanism of endothermic SF we study solid-state TIPS-tetracene[14], which has the same molecular core as tetracene but is made solution processable via the addition of triisopropylsilyl (TIPS) ethynyl side groups, see Figure 1b. As we discuss below, TIPS-tetracene possesses sharp signatures for $S_1$ and $T_1$ states, which makes clear spectral assignments much easier than in tetracene. We use ultrafast spectroscopy to show that the photoexcited population acquires TT character on sub-300 fs timescales and evolves to lose $S_1$ character on a morphology-dependent 6-20 ps timescale. The TT state is long-lived and thermally dissociates into separated $T_1$ on 10 ns timescales in disordered films (at room temperature), but surprisingly, remains bound for tens of μs in polycrystalline films (due to a low triplet-hopping rate).

Figure 1a shows the energetics of TIPS-tetracene, where the energies of $S_1$ (2.3 eV) (from UV-Vis absorption) and $T_1$ (1.20-1.30 eV) (from phosphorescence)[15], indicate that fission is similarly endothermic to tetracene. In this study we investigate two film types that differ in morphology (see SI for structural characterization). We refer to these in the text as 'disordered' and 'polycrystalline'. The UV-Vis and photoluminescence spectra of TIPS-tetracene dilute solution, disordered and polycrystalline films are shown in Figure 1c. The TIPS-tetracene chemical and crystal structure is shown in Figure 1b. For the polycrystalline films, which are composed of large (~ 100 μm diameter) domains of aligned crystallites (100 nm diameter), TIPS-tetracene does not pack in a herringbone arrangement, as found in pentacene and tetracene, nor in a 'brick-like' motif, as in TIPS-pentacene. Instead, the TIPS-tetracene molecules are arranged in loose 1D chains, where the molecules are offset with little overlap of the tetracene cores. Disordered films contain smaller aligned domains (~30 nm in diameter), of undetermined molecular packing, surrounded by large amorphous regions.

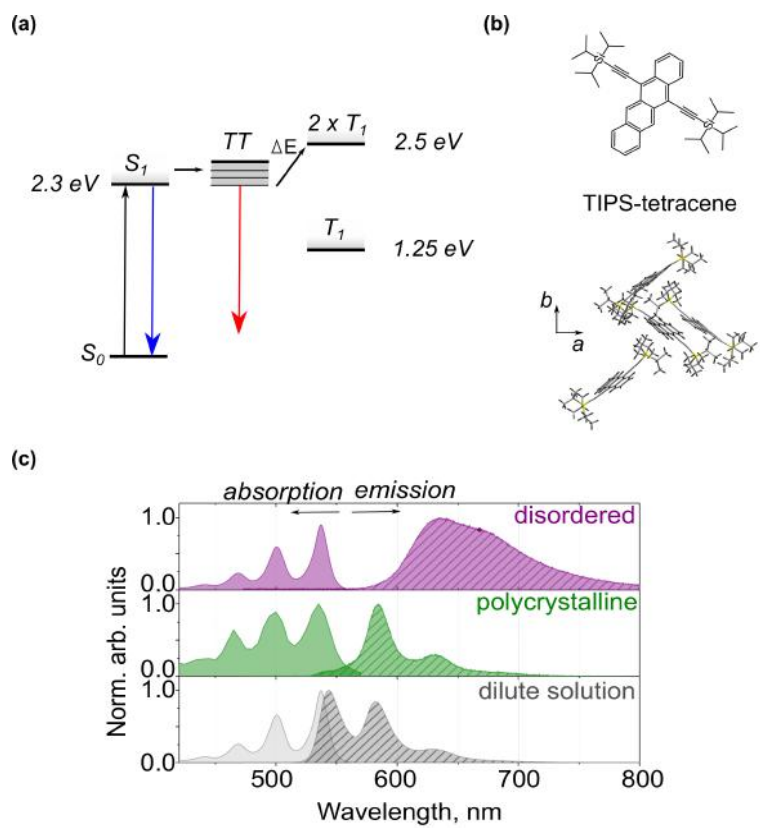

**Figure 1: TIPS-tetracene energies and structure (a) Energy level diagram of TIPS-tetracene. The $S_1$ energy is determined from the UV-vis spectra (c) whilst the $T_1$ energy is estimated from phosphorescence measurements[15]. (b) TIPS-tetracene chemical structure and one unit cell of the TIPS-tetracene crystal structure. (c) Normalized UV-vis absorption and steady-state emission of the disordered film (purple) and dilute solution (3 mg/ml in chloroform) (grey) and the excitation and emission spectra of the polycrystalline film (green).**

To investigate the dynamics of SF we use ultrafast broadband transient absorption (TA) spectroscopy with 16 fs time resolution. Figures 2a and b show the room-temperature TA spectra from 50 fs to 2 ps of the two film types, pumped with a 50 nm broad pulse centered at 530 nm (2.33 eV), close to the absorption edge. The spectral shapes observed in both films are consistent with the $S_1$ and TT species previously identified in concentrated solutions of TIPS-tetracene[15]. The initial positive signal at 570 nm (Figures 2a and b) is consistent with photoluminescence maximum of the 0-1 band of the $S_1$ emission and is assigned to $S_1$ stimulated emission (SE), as seen for dilute solution (Figure 2c). The broad negative signal from 600-1300 nm shows similarities to the solution spectrum and contains contribution from $S_1$ photo-induced absorption (PIA).

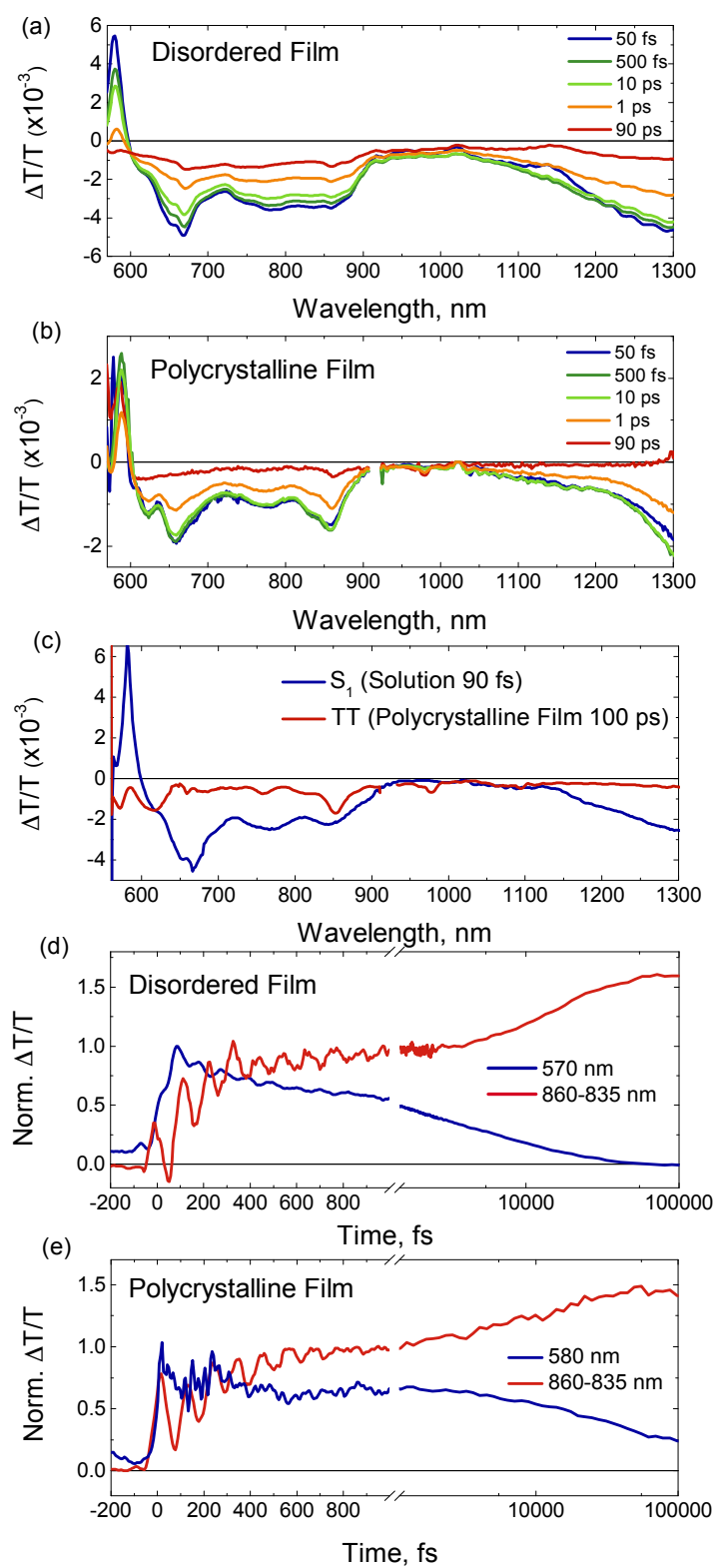

**Figure 2. Ultrafast TT formation. (a) and (b)** Ultrafast transient absorption measurements of disordered and polycrystalline TIPS-tetracene films from 50 fs – 90 ps. We note the polycrystalline film measurement is affected by pump scatter in the SE spectral region. **(c)** Transient absorption spectra of $S_1$, from a measurement of solution, and TT, from a **measurement of a polycrystalline film at a time delay of 100 ps. (c) and (d) Normalised kinetics (normalised at 2 ps) obtained from the measurements in (a) and (b) respectively. (c) For the disordered film, the kinetic representing the decay of the SE is taken at 570 nm. For the rise of the TT state we plot the difference between the change in absorption at 860 nm (a TT absorption peak) and 835 nm. The two kinetics are normalised at 0 fs and subtracted. At the beginning of the measurement both regions contain $S_1$ absorption and the difference in the kinetics represents the growth of TT. (d) For the polycrystalline film the loss of SE intensity is represented at 580 nm where there is slightly decreased pump scatter. The TT kinetic is the difference in absorption intensity between 860 nm and 835 nm, as in the disordered film.**

At longer time delays we observe a sharply peaked absorption spectrum that is assigned to the TT state, seen clearly in the timeslice at 90 ps (Figure 2a and b). Our assignment for this feature is based on the spectral assignments in the previous solution study where sensitisation measurements reveal that the $T_1$ absorption in TIPS-tetracene shows sharp peaks across the visible and NIR, regularly spaced by a vibrational frequency of ~1300 cm$^{-1}$. The TT state was shown to display the same sharp $T_1$ absorption peaks, but shifted by up to 5 meV[15]. In the film we observe similarly sharp TT absorption bands at the same position as found in solution. The assignments of $S_1$ and TT to the spectral species are based on the similarity of the spectral features we observe to those of individual $S_1$ and $T_1$ excitons, but does not preclude mixing of CT states, which our spectroscopic measurements can not give information on, into these states[13].

In Figure 2c we present the TT absorption spectrum at 100 ps obtained from a TA measurement of the polycrystalline film. We note that unavoidable pump scatter in the region of the SE in the polycrystalline film affects the spectral shape in this region (570-600 nm). Notably, the sharp TT absorption bands at 670, 850 nm and 960 nm enable us to track the conversion of $S_1$ to TT.

Figure 2d and 2e show the loss of $S_1$ SE and growth of the TT absorption for the two film types over the first 2 ps. To single out the TT growth we use kinetics at 860 nm and 835 nm, at the maximum and to the side of the sharp TT absorption band (see TT absorption in Figure 2c) and normalise to the peak initial signal when only $S_1$ is present. The difference between the two kinetics provides a background-free kinetic and captures TT population evolution. Both films show a rise time of 250 fs for the TT absorption that is matched by the rate of loss of SE intensity at 570-580 nm. We confirm this time for the ultrafast interconversion of $S_1$ and TT using a spectral deconvolution method that takes into account all spectral changes across the visible-NIR region (see SI for details). At longer times, there is a further ~10 ps rise in the TT state, matching the decay of $S_1$. We will return to this delayed rise later. The initial TT rise is orders of magnitude faster than previously considered for an endothermic fission system. But how can the supposedly higher-lying TT state be accessed on this timescale without activation?

The ultrafast pump pulse used in these experiments, which is shorter than the vibrational period of many of the modes of the molecule, generates vibrationally coherent wavepackets in both the ground and excited state potential energy surface (PES), as has been discussed previously[16]. This vibrational coherence results in strong oscillations in the TA spectra and kinetics, as seen in Figure 2d-e. We note that all absorption peaks oscillate laterally, suggesting a strong contribution from wavepackets on the excited state PES (Figure S15-16). We globally fit the population decay for the TA measurements and Fourier transform the residuals that contain the modulation on top of the electronic response (Figure S14). In Figure 3a we compare these frequencies obtained from integration across the whole spectrum of both film types (560 –1300 nm) with dilute solution (grey trace) and the ground state resonance Raman spectrum (black trace). The solution and resonance Raman spectra include low frequency modes at 315 cm$^{-1}$, 500 cm$^{-1}$ and 650 cm$^{-1}$, of which equivalent modes in tetracene have been assigned to deformation C-C-C stretches[17]. In addition, high-frequency modes are observed between 1200 and 1300 cm$^{-1}$, a region usually assigned to C-C ring stretching[17]. In the film spectra (green and purple traces) we observe additional modes that are not present in either the solution or the ground state spectra, namely modes at 760 cm$^{-1}$, 890 cm$^{-1}$ and 1090 cm$^{-1}$. This frequency range is associated with C-C deformation stretches in tetracene[17].

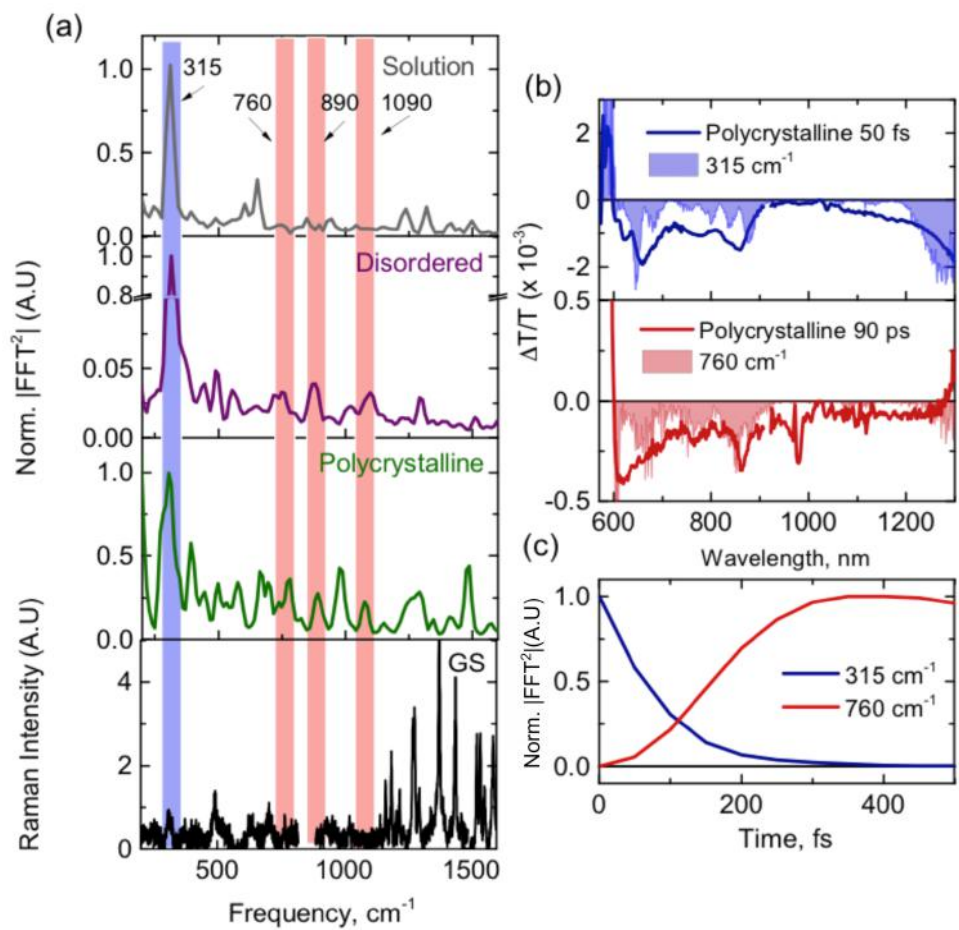

**Figure 3: Vibrationally coherent TT formation. (a)** Vibrational frequencies of TIPS-tetracene obtained from the ultrafast transient absorption measurements, from 0-2 ps, of solution (grey), disordered film (purple) and the polycrystalline film (green) and the ground state resonance Raman spectrum (black). The highlighted regions indicate a prominent $S_1$ frequency at 315 cm$^{-1}$ (blue shading) and new modes that exist in the films and not in either solution or ground state Raman at 760 cm$^{-1}$, 870 cm$^{-1}$ and 1090 cm$^{-1}$ (red shading). **(b)** The spectral slices of the 315 cm$^{-1}$ and 760 cm$^{-1}$ modes and timeslices from the transient absorption measurements of the polycrystalline film, taken at 50 fs and 90 ps. **(c)** Sliding-window Fourier transform plots of the 315 cm$^{-1}$ and 760 cm$^{-1}$ modes in the polycrystalline film, obtained at 570 nm for the 315 cm$^{-1}$ mode and 850 nm for the 760 cm$^{-1}$ mode. The sliding window Fourier transform was performed by sliding a 1 ps time window from 0 fs to 2 ps. The x axis represents the starting time for the sliding window.

The presence of these additional modes in the films may result from additional excited states present in the films that are absent in dilute solutions. To explore this, we compare the 315 cm$^{-1}$ mode, which is strongly associated with the S$_1$ state as seen in the dilute solution, with the 760 cm$^{-1}$ mode that is only present in the films. We plot the strength of these modes as a function of wavelength in Figure 3b, for polycrystalline films that afford the best signal to noise. These plots show which parts of the spectrum the modes are associated with and hence which excited states they are coupled with. As can be seen in Figure 3b, the distribution of the 315 cm$^{-1}$ mode is well matched with the spectrum of the polycrystalline film at 50 fs, when the system is dominated by S$_1$. In contrast, the distribution of the 760 cm$^{-1}$ mode closely matches the spectrum of the polycrystalline film at 90 ps, when the system is dominated by TT. This strongly suggest that the 760 cm$^{-1}$ mode is coupled to the TT state and explains its absence in the ground state Raman and the vibrational spectra of dilute solutions where SF does not occur.

In Figure 3c we present sliding-window Fourier transform plots of the 315 cm$^{-1}$ and 760 cm$^{-1}$ modes. This analysis uses a 1 ps-wide sliding window, represented on the x axis by the earliest time point in the window. The 315 cm$^{-1}$ mode is plotted for 570 nm and the 760 cm$^{-1}$ mode is plotted for 850 nm, the regions associated with SE and strong TT absorption in the polycrystalline films respectively. We find the 315 cm$^{-1}$ mode shows a decrease over time, as expected for a mode where vibrational coherence is generated upon photoexcitation and subsequently damped, either by movement away from the S$_1$ PES or by scattering on the phonon bath. Scattering is unlikely, as damping times for this phenomenon are expected to be on the picosecond timescale. The rapid damping of this mode (~500 fs) suggests rather that it is a tuning mode for the ultrafast formation of TT, indicating low frequency vibrations of the tetracene core are important in this process. In contrast, the 760 cm$^{-1}$ mode shows an initial increase, as the window slides from 0-1 ps to 0.4-1.4 ps. The time period of the 760 cm$^{-1}$ mode, 43 fs, is much shorter than the TT rise time (250 fs), which means that vibrational coherence cannot be created impulsively via the SF process populating TT. Rather, the 760 cm$^{-1}$ mode is likely to be a product mode of the reaction, formed as the nuclear wavepacket crosses from the S$_1$ to the TT PES. Hence, the increase in strength seen in Figure 3c as the underlying TT state grows in (Figure 2e) indicates the formation of the TT state occurs via a vibrationally coherent process. This is similar to vibrationally coherent SF observed in exothermic SF systems, which has been explored both theoretically[18,19], and experimentally[20,21,22], and also vibrationally coherent ultrafast internal conversion in polyenes[23] and energy transfer in biological light harvesting systems[24]. The observation of vibrationally coherent formation of TT states also implies a breakdown of the Born–Oppenheimer approximation and strong coupling of nuclear and electronic degrees of freedom, which in its limit give rise to a conical intersection.

This coupling also modulates the energy levels of the states involved in the SF process, which is likely to help drive TT formation, as has been previously suggested in a computational studies of tetracene derivatives[25]. The lateral shifting of the PIA features in the ultrafast TA spectra reveal that the transition energies over all excited states present are modulated by more than 100 meV (Figure S14), indicating that energies calculated with the ground state geometries do not provide a good guide to understanding what happens on the excited state PES.

Our observations show that vibrational modes are involved in the rapid formation of TT in TIPS-tetracene. This is different from a model based on strong-electronic coupling between $S_1$ and TT, which has previously been invoked to explain the dynamics of tetracene by Zhu *et al.*[26]. The lack of a shift in absorption from solution to solid-state indicates weak interactions between adjacent molecules in TIPS-tetracene in the ground state, which would be inconsistent with a strong-electronic coupling model. In contrast, as has been shown for pentacene, vibronic coupling allows mixing of states even when the direct electronic coupling of $S_1$ and TT is low (<200 cm$^{-1}$)[20]. Instead of the formation of a superposition of $S_1$ and TT upon photoexcitation, our data suggest that following photoexciation vibrational modes drive the wavepacket from the initially populated Frank-Condon region towards the state crossing where $S_1$ and TT PES are linked, thus forming a single multi-dimensional PES. While such a process is known for exothermic fission, where TT is lower in energy than $S_1$, these results show how even in endothermic systems the TT state can be accessed on ultrafast timescales, enabling endothermic fission to be equally efficient at generating triplet excitons.

As seen in Figure 2d and 2e, following the early time ultrafast conversion of $S_1$ to TT, there is delayed rise in TT with a concomitant decay of $S_1$. This slower $S_1$ decay is reminiscent of the $S_1$ decay observed in tetracene[6,10] – it decays independently of temperature over tens of picoseconds (see Figure S18-19) and shows a morphology-dependent lifetime (6-10 ps for disordered and 15-19 ps for polycrystalline), matching the delayed rise of TT. Correspondingly, a 11 ± 1 ps and 12 ± 1 ps photoluminescence lifetime was measured for the disordered and polycrystalline film types respectively using a transient grating photoluminescence set-up with a time resolution of 200 fs (TGPL)[27] (Figure S10). In TA, for the polycrystalline samples we note a variation in lifetimes by ~1-2 ps between different positions in the film. Thus, the variation in lifetime for the $S_1$ decay we observe between the TA and PL, and within the TA measurements, we consider to arise due to the inhomogeneity across the polycrystalline film. Importantly, in the absence of higher time resolution and sharp TT spectral signatures this $S_1$ decay time would appear to be the SF rate for TIPS-tetracene. However, as we have shown, this $S_1$ decay rate gives a misleadingly slow indication of the rate of initial TT formation.

We consider that the ultrafast formation of TT represents SF that occurs at photoexcited sites in the film where intermolecular arrangement is optimal for SF. We find that the prompt vs delayed TT formation ratio is larger in the disordered film. The ultrafast rise of TT makes up about 70% of the final TT strength in the disordered film, compared to 30 % for the polycrystalline film. The picosecond morphology-dependent loss of $S_1$ and rise of TT may to be related to the time needed for the excitation to diffuse to optimal fission geometries, as has been suggested in tetracene[10,28]. Multiple timescales for TT formation have also been observed in hexacene, where fission is exothermic, and have been attributed to both coherent and incoherent fission processes[22]. However, we cannot rule out an alternate explanation for the slow loss of $S_1$ that is also consistent with our data, whereby vibronic coupling sets up an equilibrium between $S_1$ and TT and the 6-19 ps timescale is related to the time needed to fully shift the $S_1$-TT equilibrium to only TT. Such a process could be mediated by slow damping of certain low-energy phonon modes that are associated with the photoexcited state and not present within the crystal in the ground state, and thus cannot be easily damped. This could allow for a vibronic equilibrium between $S_1$ and TT to be setup, which would last as long as the modes are undamped. Damping of vibrational modes over 10s of ps has been previously observed at low temperature in pentacene crystals in naphthalene[29]. This hypothesis would be supported by our observation that the broad PIA (600 nm-1300 nm) in the two films does not show the prompt loss of oscillator strength as seen the SE feature, but shows many sharp features associated with TT from sub-200fs timescales and decays over the picosecond timescale (Figure 2a and b). We predict that the ultrafast TT formation is associated with rapid movement away from the Franck Condon region resulting in loss of SE and that the broad PIA features are due to absorption of the resulting $S_1$-TT state to higher-lying excited states.

The fast formation dynamics indicate that the initial TT yield could be very high, as it outcompetes radiative and non-radiative decay channels. However, it is the yield of $T_1$ + $T_1$ at longer timescales that is more relevant for photovoltaic device applications. Figure 4e tracks the evolution of both films at 850 nm from 10 ps to 2 ms. In the disordered film, the raw TA kinetic shows the presence of two decay regimes that we can spectrally resolve, using a spectral deconvolution code based on a genetic algorithm[29], into the decay of bound TT state ($\tau$= 10 ns (295 K)) and the decay of separated $T_1$ ($\tau$= 10 µs (295 K)) (Figure 4b). The near-IR spectra of the concentrated solution, disordered film and polycrystalline film are shown in Figure 4(a,c,e). For the concentrated solution and disordered film we observe a shifting of the TT absorption peaks and loss of absorption between the two peaks over 1 ns - 1 µs, to give the absorption confirmed via sensitisation to be $T_1$ at microsecond delays[15]. A shift in the TT absorption peaks have also been observed in pentacene derivatives and associated with the changing excitonic interactions of the bound state as it separates[31]. However, in the disordered film of TIPS-tetracene the TT dissociation rate and the yield of $T_1 + T_1$ are both temperature-dependent (Figure 4b). As the temperature is lowered we observe slower TT decay and relatively weaker $T_1 + T_1$ absorption, consistent with a thermally activated TT separation. At the lowest temperatures measured (10 K), we resolve only the TT state. We note that the room temperature TT lifetime measured here is comparable to the TT lifetime in concentrated solution (8.7 ns)[15], indicating that a similar barrier is overcome in both systems. We make two estimates for the yield of the reaction TT to $T_1 + T_1$. From the quenching of the TT PL at room temperature (see below), we obtain a value of 180 ± 10 %, and from the evolution of the TA, using $T_1$ cross-sections obtained from sensitization measurements a value of 130 ± 20 % (see SI).

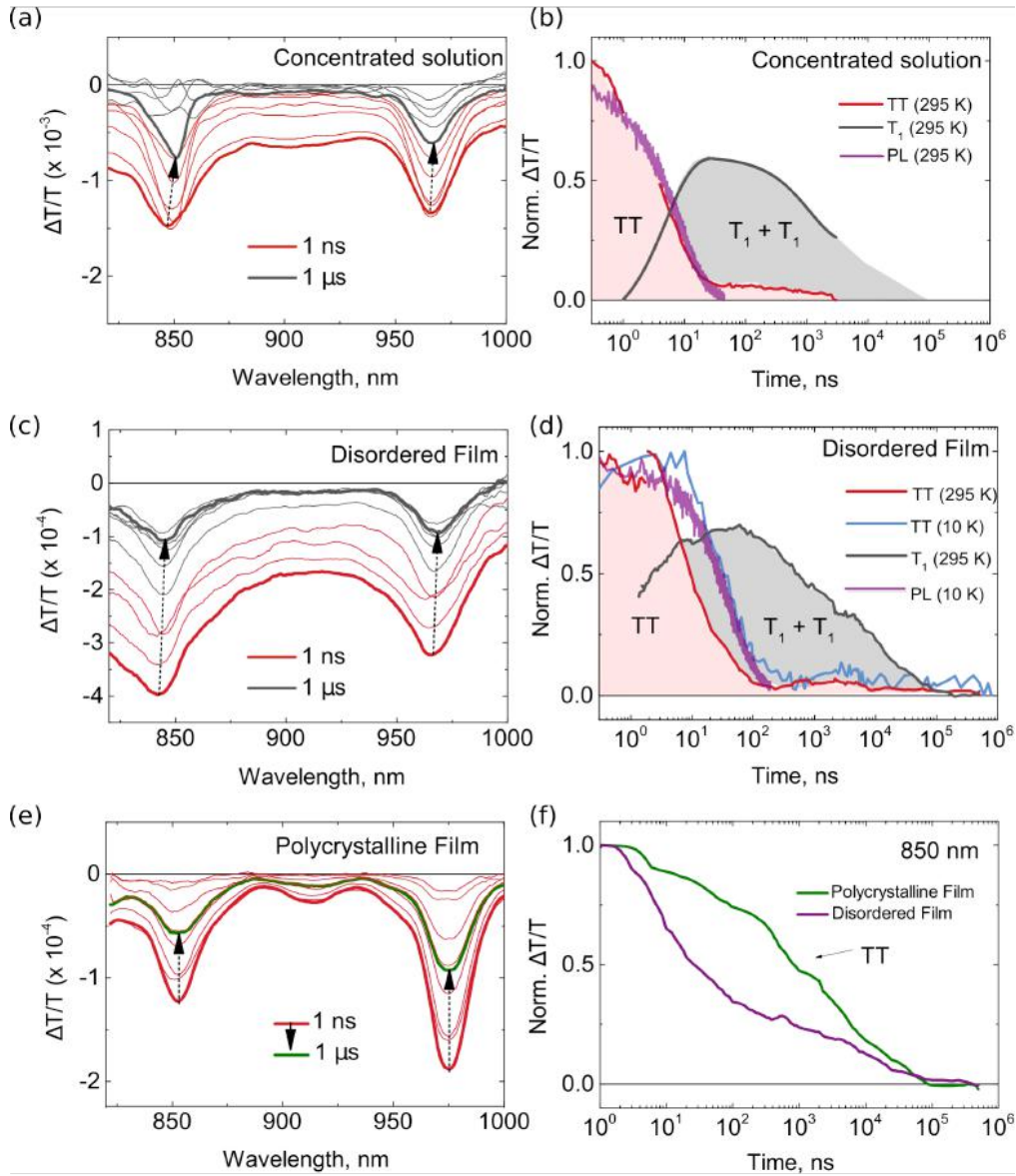

**Figure 4. Thermally-activated TT separation. (a,c,e) Transient absorption spectra from 1 ns to 3 μs of the (a) concentrated solution, (c) disordered film and (e) polycrystalline film in the near-IR spectral region. For the concentrated solution and disordered film we observe two decay regimes that can be spectrally deconvoluted using a code based on a genetic algorithm (see SI for details) into two distinct spectra corresponding to TT and $T_1 + T_1$. The TT spectrum contains additional absorption intensity between the two peaks at 850 nm and 960 nm and the peaks are blue-shifted by 5 meV. (b,d) The extracted room temperature kinetics for the TT and $T_1$ decay in the (b) concentrated solution and (d) the disordered film. The PL decay of the TT state at room temperature in the concentrated solution and at 10 K in the disordered film are shown plotted against the TA decay (purple trace). (f) Raw kinetics from the transient absorption measurement of the disordered and polycrystalline films at 850 nm highlighting the different decay behaviour.**

The TT absorption, including the weaker TT feature at 900 nm, decays more slowly and is present over the full decay in the polycrystalline film, Figure 4e. Furthermore, the spectrum cannot be de-convoluted into multiple species. The decay of the spectrum does not show an evident temperature dependence, nor is its decay accelerated under increased fluence (from 80-400 µJcm$^{-1}$). Taken together, these observations suggest that in polycrystalline films the TT state does not dissociate to $T_1 + T_1$ even at room temperature.

To understand why the crystalline morphology gives rise to a long-lived TT state, we calculate the triplet-hopping rates in two distinct intermolecular situations that represent the extreme case for structure difference between the polycrystalline and disordered films - two molecules from the crystal structure and two pi-pi stacked dimers, respectively. We chose a pi-stacked dimer to represent fission sites in the disordered film because the TT PL signatures (introduced below) suggest the TT state occupies two molecules in a low-energy optimal pi-stacked geometry as found in solution. Using the linear response time-dependent density functional theory (TDDFT) with the fragment excitation difference (FED) method, we find that the two intermolecular geometries present very different hopping integrals: 15 and 0.2 meV for the pi-stacked dimer and the crystal structure respectively. Using a reorganization energy of 0.33 eV and the Marcus model[32] these hopping integrals correspond to a $T_1$ hopping time of 2.5 ps and 52 ns, respectively. As we cannot know how large or dense the pi-stacked domains are in the disordered film, this value represents an upper limit. The true hopping rate may be lower than this value. In addition, we note that these estimated hopping times assume an ideal environment at 0 K, and as thermal fluctuations in the triplet energies and $T_1$ hopping couplings are included these times are expected to get slower at higher temperatures[32]. Importantly, these calculations highlight the unusually poor triplet-triplet coupling in the TIPS-tetracene crystal structure. While the chosen structures represent the largest possible difference between the polycrystalline and disordered films, the large variation in hopping rates suggest that in polycrystalline films, triplet hopping is very slow and significantly slow the dissociation of the bound TT state.

Photoluminescence measurements also reveal information on the energetics and evolution of the TT state. The photoluminescence quantum efficiency in both film types is moderately low: 3% and 1% for the disordered and polycrystalline films respectively, at room temperature. However, as we report below, the PL yield for the disordered film increases rapidly with reducing temperature, by a factor of 20 at 10 K, implying a PL yield of around 60%. Fig. 5c and 5f show the temperature-dependent steady-state PL of the two films. The disordered film shows a broad PL centered at ~650 nm, with two vibronic peaks clearly visible at low temperature. The polycrystalline film shows three different peaks, and the overall temperature-dependent behavior is similar to tetracene[5,6], with the second peak at 580 nm dominating at higher temperatures and the high-energy feature at 540 nm, associated with the 0-0 transition of $S_1$ dominating at low temperature.

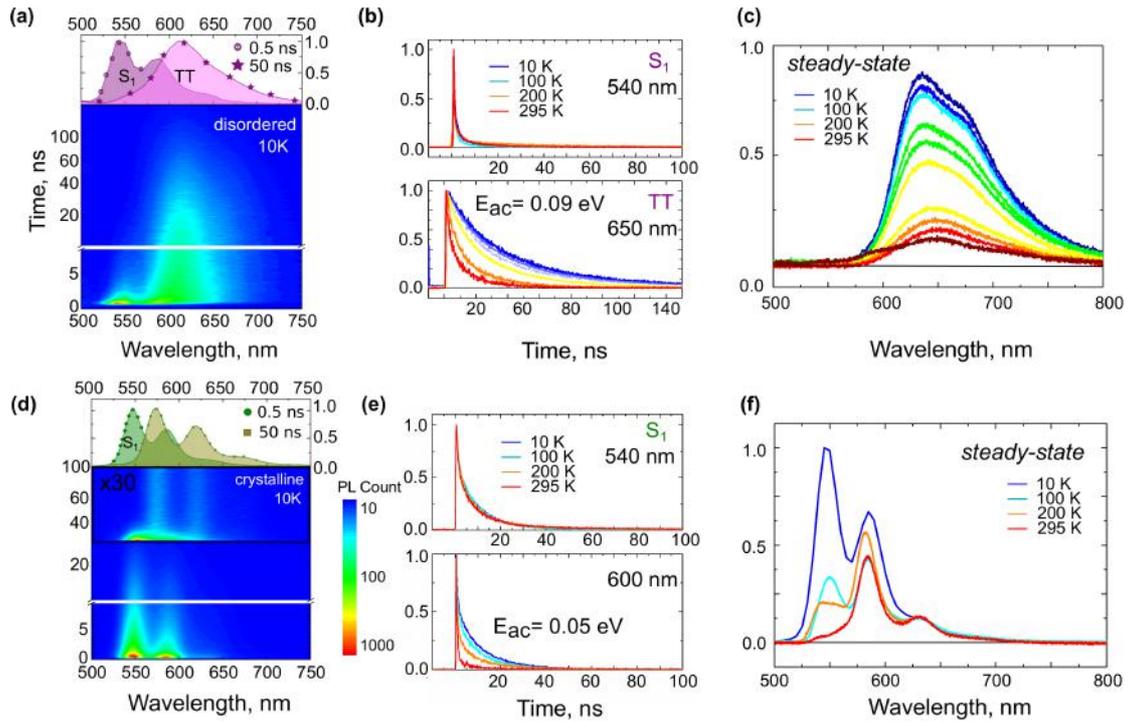

**Figure 5: Thermally-activated TT emission.** (a,d) Time-resolved emission scans reveal two emissive states in both the (a) disordered film and (d) polycrystalline films at 10 K. The emission from 30-100 ns is shown magnified for the crystalline film (inset (d)). Timeslices at 0.5 ns and 50 ns are shown above the colour plots. (b,e) Temperature-dependent kinetics taken at 540 nm and 600 nm for the (b) disordered and (e) polycrystalline films. We obtained an activation energy in the red-shifted region of ~90 meV and ~50 meV respectively. (c,f) Temperature-dependent steady-state emission for the (c) disordered and (f) polycrystalline films. The spectra are normalised to the peak of the emission at 10 K. At higher temperatures the spectra are plotted to show their intensity relative to the 10 K emission.

The time-resolved PL at 10 K reveals that in both film types there are two emissive species, a prompt high-energy component, and a red-shifted emissive state (time slices at 0.5 ns and 50 ns in Figure 5a and 5d). The high-energy state has the same spectrum as the dilute solution and decays with a temperature-independent lifetime (Figure 5b and 5e); we assign this to emission from $S_1$ in both disordered and polycrystalline films. For the crystalline film, the $S_1$ feature has a delayed component, similar to the well-studied delayed PL in tetracene[6,33]. For the disordered film, no delayed component of $S_1$ is detected, indicating that any regeneration of the $S_1$ from TT is too weak to detect.

In both films the red-shifted emission region shows a thermal-dependence. From the kinetic traces at 650 nm of the disordered film we obtain an activation energy, out of the red-shifted emissive state, of ~90 meV (Figure S8). This temperature-dependent emission tracks the temperature dependence for the TT state we observe in TA (Figure 4b).

The red-shifted emission in the polycrystalline film is weak (orders of magnitude weaker than the prompt $S_1$ at 10 K), making it difficult to difficult to assign this emission to a particular state. However, in the disordered film the red-shifted emission is strong and can be compared to the excited states identified in TA. We note that the high PL yield from the red-shifted emission in the disordered film at 10 K allows direct comparison of the PL decay with the time evolution of the TA spectra. As shown in Figure 4d the PL decay occurs on a 50 ns timescale and matches the TT state decay measured via TA, as seen in solution (Figure 4b). We thus consider that the red-shifted emission in the disordered film also derives from the TT state. We note that 50 ns lifetime with 60% PL yield at 10K suggests a radiative lifetime for the TT state in the disordered film of 80 ns. From the PL yield in the disordered film, we can estimate the yield of the reaction TT to $T_1 + T_1$ to be ~180%.

For both films, the red-shifted species shows pronounced vibronic structure. These are sharper in the polycrystalline material and broadened in the disordered, where the emission has the same spectrum as the TT emission in concentrated solution (Fig. S14). We therefore predict that the weak, red-shifted emission in the polycrystalline film is also due to the TT state and that the transition dipole moment of the TT state in the polycrystalline film is significantly reduced compared to the disordered material. This is likely to be due to the ability of the molecules in the disordered film to rearrange into an excimer-like TT state geometry that can more easily radiatively couple to the ground state.

The activation energy we extract for TT dissociation from the PL measurements (50-90 meV), is comparable to the activation energies measured for tetracene using PL and TA (40-70 meV)[33,34]. Taking into account the entropic gain following the dissociation of TT to free triplets, which has been discussed by others[25,35], we expect the activation energy to be one half of the value of $E_b$ which we estimate to be 200 meV.

In summary, we have tracked the photophysical behavior of this system over ten orders of magnitude (Figure 6) and our results show that ultrafast, activation-less formation of stabilized, long-lived TT states, that quench radiative losses via $S_1$ and protect the excitation from competing decay channels, is key to efficient endothermic formation of $T_1 + T_1$. These states can be formed on sub 300 fs timescales, the ultrafast conversion mediated by vibronic coupling. These results unify the observation in tetracene of the simultaneous ultrafast rise of $S_1$ and multiexciton features by Zhu *et al.* by photoelectron spectroscopy[26] with the slower $S_1$ dynamics observed in optical measurements[6,10,12]. At later times, the long-lived TT states can be thermally dissociated to free $T_1$, if the crystal morphology supports efficient triplet hopping. If not, TT can remain bound on µs timescales without being dissociated to free T$_1$. More generally, our results demonstrate how vibronic coupling and structural relaxation can allow for the formation excited states on sub-ps timescales, that are subsequently stabilised and long-lived allowing sufficient time for endothermic process to occur efficiently at longer timescales. These results pave the way for further studies of how the optimisation of the chemical structure of endothermic fission materials can alter vibronic coupling and the ultrafast fission process.

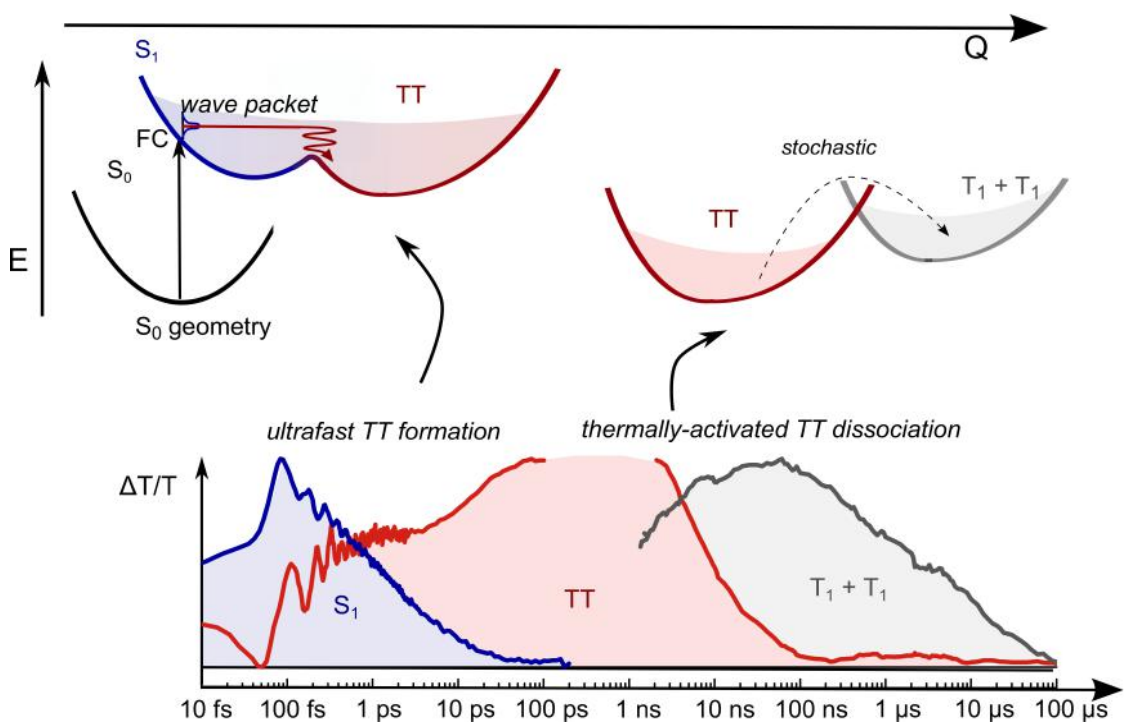

**Figure 6: The role of the TT state in endothermic singlet exciton fission.**
**(Top) A schematic diagram of endothermic singlet exciton fission. The wavepacket generated in S$_1$ is passed to the TT state during ultrafast formation of TT. Thermally-activated TT separation occurs over tens of nanoseconds. (Bottom) The kinetics of the S$_1$, TT and T$_1$ + T$_1$ species in the disordered film from 10 fs to 100 µs, obtained from the TA measurements using a genetic algorithm for spectral deconvolution.**

**Author Contributions**

H.L.S and A.C carried out the experiments, interpreted the data and wrote the manuscript. SRY and MHG ran calculations, interpreted the data and wrote the manuscript. NCG and SB interpreted the data. KB, AC, KC and JH performed experiments. KT and JA designed and synthesized the materials. AR, AJM and RHF interpreted the data and wrote the manuscript.

**Acknowledgments**

The authors thank the Winton Programme for the Physics of Sustainability and the EPSRC for funding. RHF thanks the Miller Institute for Basic Research and the Heising-Simons Foundation at UC Berkeley for support. The authors thank Dr. T. Arnold (Diamond Light Source), Dr. J. Novak, D. Harkin and J. Rozboril for support during the beamtime at beamline I07 and the Diamond Light Source for financial support. The computational work was supported by the Scientific Discovery through Advanced Computing (SciDAC) program funded by the U.S. Department of Energy, Office of Science, Advanced Scientific Computing Research, and Basic Energy Sciences.

Supporting Material

# Ultrafast Triplet Pair Formation and Subsequent Thermally-Activated Dissociation Control Efficient Endothermic Singlet Exciton Fission

Hannah L. Stern[1], Alexandre Cheminal[1], Shane R. Yost[2,3], Katharina Broch[1], Sam L. Bayliss[1], , Kai Chen[4,5], Maxim Tabachnyk[1], Karl Thorley[6], Neil Greenham[1], Justin Hodgkiss[4,5], John Anthony[6], Martin Head-Gordon[2,3], Andrew J. Musser[1], Akshay Rao[1] and Richard H. Friend[1].

[1] Cavendish Laboratory, University of Cambridge, UK

[2] Kenneth S. Pitzer Center for Theoretical Chemistry, Department of Chemistry, University of California, Berkeley, USA.

[3] Chemical Science Division, Lawrence Berkeley National Laboratory, Berkeley, USA.

[4] MacDiarmid Institute for Advanced Materials and Nanotechnology, New Zealand.

[5] School of Chemical and Physical Sciences, Victoria University of Wellington, New Zealand.

[6] University of Kentucky, Lexington, USA

## *Contents*

## 1.) Methods

### a.) Materials

TIPS-tetracene was synthesized according to the procedure in reference [1]. This material incorporates two triisopropylsilylethynyl (TIPS) groups coordinated to the para-positions of the second benzene ring in tetracene. The TIPS side groups make TIPS-tetracene highly soluble (up to 300 mg/ml in chloroform) and substantially change the crystal structure (see structural characterization below).

For all of the optical measurements TIPS-tetracene was either spin coated or drop cast onto 13 mm diameter fused silica substrates in an oxygen free environment. Samples were measured under vacuum or, for low temperature measurements, in a helium dynamic flow cryostat.

### b.) Optical Spectroscopy

UV-Vis absorption spectra were measured on a Cary 400 UV-Visible Spectrometer over the photon energy range 1.55 eV-3.54 eV. Steady-state photoluminescence spectra were collected using a pulsed laser at 2.64 eV (PicoQuant LDH400 40 MHz) and collected on a 500 mm focal length spectrograph (Princeton Instruments, SpectraPro2500i) with a cooled CCD camera.

Time-resolved photoluminescence decay was measured using time-correlated single photon counting (TCSPC), an intensified CCD camera (ICCD) and a transient grating set-up (TGPL). For all measurements the sample was measured in either a side-on or backward reflection geometry, to mitigate self-absorption. The TCSPC set-up uses the same excitation source and camera as the steady-state PL and has a temporal resolution of 300 ps.

Transient grating measurements were measured by a home-built transient-grating photoluminescence spectroscopy (TGPLS) [2]. Briefly, a Ti-Sapphire amplifier provided 470 nm excitation pulses from an optical parametric amplifier and 800 nm fundamental outputs for transient grating optical gate. The excitation intensity is ~40 $\mu J/cm^2$. No photodegradation was found during the measurement. All reflective optics was used between the sample and nonlinear gate medium to minimize the dispersion of PL. The broadband PL emission was diffracted by transient grating optical gate generated by inference of two gate beams on a 1 mm fused silica plate. The gated PL was spectrally resolved and detected by a spectrograph (SP2300 by Princeton Instruments) and an intensified CCD (ICCD, PIMAX 3 by Princeton Instrument). Spectral bandwidth of the setup is from 500 to 740 nm limited by bandpass filters to block the residual of 470 nm excitation and 800 nm gate laser pulses. The electronic gate width of the ICCD was set at 10 ns. It took 10 second for the ICCD to accumulate 30000 shots for a time-resolved spectrum at each time delay. Four scans were averaged to improve the signal-to-noise ratio of time-resolved spectra and kinetics.

Nanosecond- millisecond spectrally-resolved photoluminescence measurements were made using an intensified CCD camera (ICCD). For the ICCD measurements the excitation source was the output of a home built NOPA, a 2.33 eV laser pulse at a 1kHz repetition rate.

The photoluminescence quantum efficiencies of the films were measured using an integrating sphere and a 2.33 eV excitation source. These measurements were performed on encapsulated films at room temperature.

Transient absorption spectra were recorded on a setup that has been previously reported [3]. Spectra were recorded over ultrafast (20 fs- 2 ps), short (50 fs- 2 ns) and long (1 ns-1 ms) time delays with probe ranges covering from 2.48 eV- 1.55 eV and 1.55 eV- 1.13 eV.

The ultra-fast (20 fs) TA experiments were performed using a Nd :YAG based amplified system (PHAROS, Light Conversion) providing 14.5W at 1025nm and 38kHz repetition rate. The TA setup is similar to the TA set-up described above, the probe beam being generated by focusing a portion of the fundamental in a 4mm YAG substrate. The pump beam is generated using a NOPA (37° cut BBO, type I, 5° external angle) pumped with the third harmonic of the source (HIRO, Light Conversion). The amplified 3mm YAG-generated white light results in 525nm centered, 25nm bandwidth 600μW pulses. The pump pulses are compressed using chirped mirrors and leads to pulse durations in the 40-20fs range. The 525nm NOPA pump pulses and white light are overlapped in the sample with a 8.7° angle using reflecting optics. The white light is delayed using a computer-controlled piezoelectric translation stage, and a sequence of probe pulses with and without pump is generated using a chopper wheel on the pump beam. After the sample, the probe pulses are seeded through a monochromator and imaged using an InGaAs photodiode array camera (Sensors Unlimited/BF Goodrich) so as to measure the transient difference absorption between 540 and 1000nm at a 19kHZ rate.

For short-time (ps-TA) measurements a portion of the 1 kHz pulses from the central Ti:sapphire amplifier system (Spectra-Physics Solstice) was feed into a TOPAS optical parametric amplifier (Light Conversion) to produce a tunable narrowband 2.33 eV pump beam. A second portion is directed into a series of home-built NOPAs, modelled on Cerullo *et al.*[4], to generate probe beams in the visible and near-IR. For short-time measurements the probe is delayed using a mechanical delay-stage (Newport). For long-time (ns-TA) measurements a separate frequency-doubled Q-switched Nd:YVO4 laser (AOT-YVO-25QSPX, Advanced Optical Technologies) is used to generate the pump. This laser produces pulses with a temporal breadth below 1 ns at 2.33 eV and has an electronically controlled delay. The pump and probe beams are overlapped on the sample adjacent to a reference probe beam. This reference is used to account for any shot-to- shot variation in transmission. The beams are focused into an imaging spectrometer (Andor, Shamrock SR 303i) and detected using a pair of linear image sensors (Hamamatsu, G11608) driven and read out at the full laser repetition rate by a custom-built board from Stresing Entwicklungsburo.

Initial measurements were recorded at a range of laser fluences (5-400 μJ/cm$^2$) on both the ps and ns transient absorption set-ups. We find $S_1$ decay is unaffected by $S_1$-$S_1$ annihilation up to 100 μJ/cm$^2$. The same rate of decay across the fluence series rule out any bi-molecular decay below 100 μJ/cm$^2$. For the measurements on the crystalline films we found higher pump power were needed, due to the additional noise in the measurement from scattering. We found that that the decay of the long-lived triplet absorption features are insensitive to fluence, up to 400 μJ/cm$^2$.

In all measurements every second pump shot is omitted, either electronically for long-time measurements or using a mechanical chopper for short- time measurements. The fractional differential transmission (ΔT/T) of the probe is calculated for each data point once 500 shots have been collected. Positive signals are due to stimulated emission of excited states (SE). Photoinduced absorptions of excited states (PIA) are observed at lower energies as negative signals.

## 2.) Structural Characterization

### a.) GIWAXS images

Grazing incidence wide angle X-ray scattering measurements were performed at beamline I07, Diamond Light Source, UK, using a Pilatus 1M detector and beam energy of 12.5keV.

We measured the GIWAXS images for the spin-coated disordered and the dropcast polycrystalline film. The disordered films presented in the paper are all spun from 10-30 mg/ml solution. The polycrystalline films presented in the main text are made via dropcasting from toluene solution.

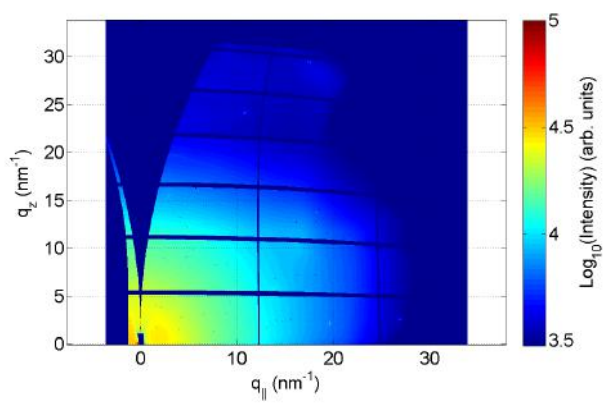

*Figure S1: GIWAXS images for a spin coated TIPS-tetracene film, spun from 10 mg/ml chloroform solution.*

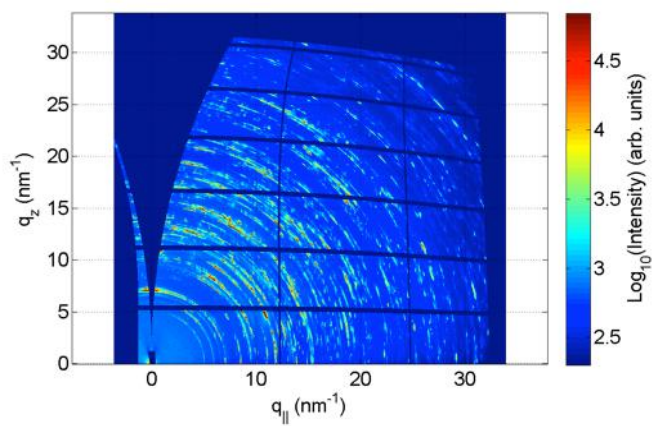

*Figure S2: GIWAXS images for a dropcast polycrystalline TIPS-tetracene film.*

For the dropcast sample we observe Debye-Scherrer rings that indicate there are crystallites with a preferred orientation. The thickness of the coherently scattering part of the film (crystallite size) was determined by the Scherrer formula, which assumes spherical grains [5]. The Scherrer formula predicts domains on the order of 0.09 μm for the dropcast sample. The spin-coated sample shows no rings present. For a film spun from 30 mg/ml, the Scherrer formula gives a crystallite size of 34 nm.

### b.) X-ray diffraction data for single crystal TIPS-tetracene and dropcast polycrystalline films.

X-ray diffraction measurements were performed using a Bruker D8 setup and a wavelength of 1.5406 angstrom. We compare the X-ray diffraction recorded for three polycrystalline films with the diffractogram of the single crystal.

Figure S3 shows the different dropcast samples all give diffraction peaks that are observed in the powder diffractogram of the single crystal. We can see the single crystal powder diffractogram displays many diffraction peaks. This is because the powder is composed of crystallites at random orientations on the film. The dropcast films show fewer peaks, indicating the measurement is sensitive to certain crystallites. This is evidence that the crystallites are packed with a preferential orientation relative to the substrate surface. We see intense peaks at 0.7 and 1.0 A-1. These peaks are assigned to the (111) and (121) planes. The (111) plane lies parallel to the substrate surface. Thus, we can determine the relative orientation of the majority of TIPS-tetracene molecules to the substrate surface- this is shown in Figure S4. The long axis of the molecule lies along the substrate surface.

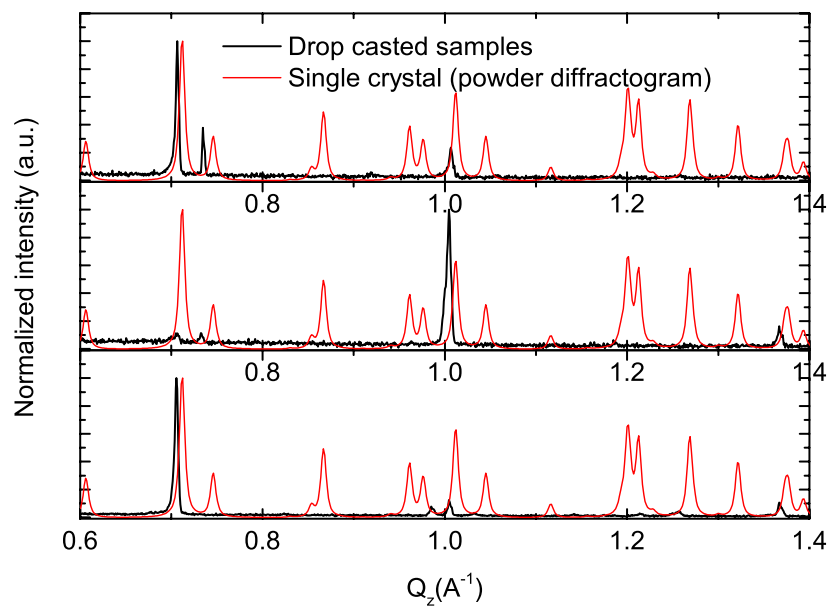

*Figure S3: X-Ray diffraction recorded for three different TIPS-tetracene crystalline films (black traces). The powder diffractogram of the single crystal was calculated based on the unit cell parameters of TIPS-tetracene using Mercury. The Bragg peaks observed for the drop cast samples have a comparable width of the diffraction peaks for the single crystal but exhibit a consistent shift to smaller $Q_z$ values which we relate to an increase in the lattice dimensions in the dropcast samples. We note that*

*not all peaks of the powder data are observable in the XRD-data of the dropcast samples due to a preferred orientation of the crystallites in the dropcast films.*

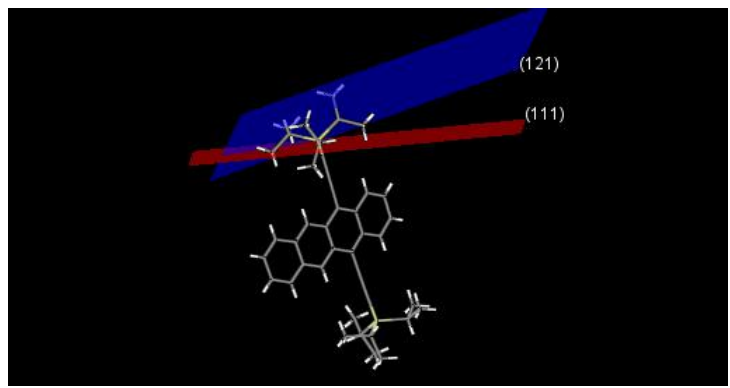

*Figure S4: Orientation of one molecule in the TIPS-tetracene structure with respect to the (111) and (121) crystal planes.*

### c.) Determination of average crystallite grain size.

The average crystallite size of the dropcast films was estimated from the full width half maxima of the Bragg-peaks in the XRD data using the Scherrer formula. This method assumes spherical grains, thus is expected to give just a rough estimate. From our collection of 13 dropcast samples we determined an average crystallite diameter of ~90 nm and a spread of sizes of ~40- 200 nm. We determine from the GIWAXS images that these crystallites are preferentially aligned and we observe in the polarized microscope images that the aligned crystallites form domains on the order of 100 μm in diameter.

### d.) Crystal structure of TIPS-tetracene.

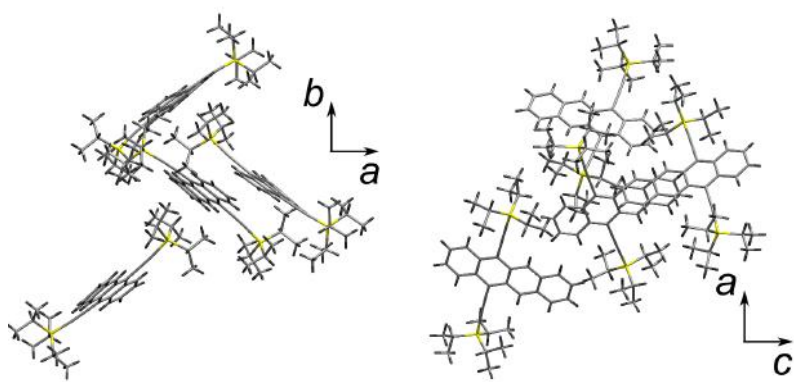

*Figure S5: One unit cell of the crystal structure of TIPS-tetracene. The unit cell is orthorhombic with a = 14.22, b = 15.13, c = 16.85 angstrom and consists of four rotationally inequivalent molecules per unit cell.*

### 3.) Photoluminescence

#### a.) Photoluminescence Quantum Efficiency (PLQE)

| Sample | PLQE |
|---|---|
| Disordered film | 0.03 ± 0.005 |
| Polycrystalline film | 0.01 ± 0.005 |
| Concentrated solution* | 0.02 ± 0.005 |
| Dilute solution* | 0.82 ± 0.01 |

*Table 1: Room temperature PLQE values for the disordered and polycrystalline films and concentrated and dilute solution.*
*\* Solutions were in chloroform. Concentrated= 300 mg/ml and dilute= 3mg/ml.*

The PLQE values at room temperature for both films is low: ~ 3 % and 1 % for the disordered and polycrystalline films respectively. As the temperature is lowered to 10 K we observe a rapid increase in PL intensity for the disordered film (x20) but less of an increase in the polycrystalline film (~ x2.5). Below, we represent the changes in intensity and spectrum with temperature, and show the time-resolved PL behavior.

#### b.) SVD Components of Time-resolved Emission Scans

We observe prompt $S_1$ emission and red-shifted TT emission in both film types. Below we display the spectra identified using singular value decomposition and their associated kinetics, in the disordered film.

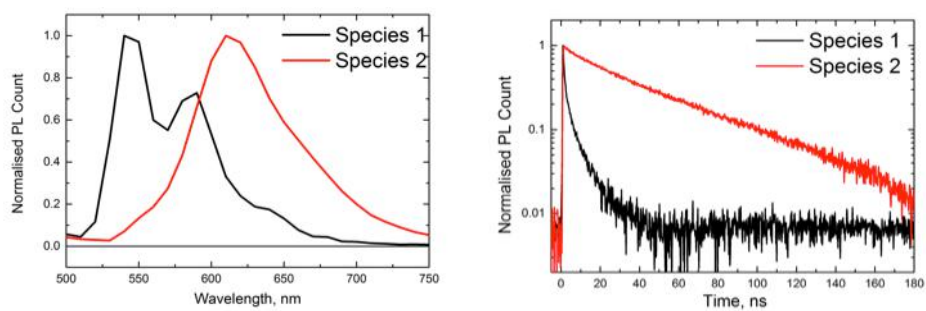

*Figure S6: We use singular value decomposition (SVD) to determine the principal components in the TRES scans. Here, we present the two main SVD components for the disordered film measurement at 10 K. We observe $S_1$ emission that is instrument limited (<300 ps) and longer-lived red-shifted emission ($\tau$= 50 ns). We note the spectral resolution for this measurement is 10 nm therefore the weak vibronic features in the TT spectrum (seen clearly in the steady-state measurement) are harder to see.*

### c.) Solution and Film TT emission

We plot the normalized steady-state emission for the concentrated solution and disordered film to highlight the similarity in spectral shape in the TT emission. The solution spectrum shows a more $S_1$ emission due to the slower $S_1$ decay rate in the solution system. The TT state in both systems shows a similar lifetime (~7.8 ns in solution and 10 ns in the film).

The similarity in spectral shape between the TT state in the disordered film and solution suggests that the molecules in the TT states find a similar intermolecular geometry in both systems. We expect that the TT excimer state formed in solution represents the optimal intermolecular interaction, a pi-stacked dimer, and we predict that such sites also exist in the disordered morphology and are preferentially populated by the TT state.

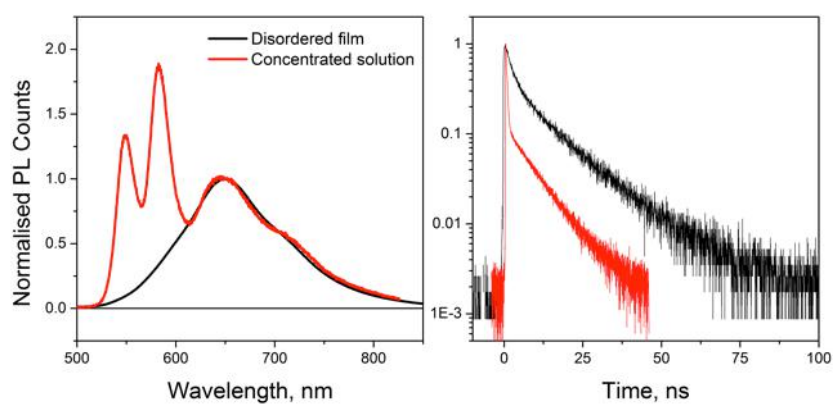

*Figure S7: Normalized steady state emission and time resolved emission (at 650 nm) of concentrated TIPS-tetracene solution (red) and the disordered film (black). The steady state emission of the concentrated solution and disordered film is normalized to the peak of the TT emission at 650 nm*

### d.) Determination of activation energy

We fit the rate of the red-shifted emission for the disordered and crystalline films against temperature, to the expression,

$$y = A + B \times e^{(\frac{Eac}{x})}$$

where,
A = rate of emission at negligible temperature.
$E_{ac}$ = Activation energy
For each film the emission displays two time constants, and here we plot the second, longer emissive decay rate against temperature. We extract an activation energy of ~90 meV for the disordered film and ~50 meV for the crystalline film. For the crystalline film there are fewer data points, and the emission is very weak, making this a rough estimate.

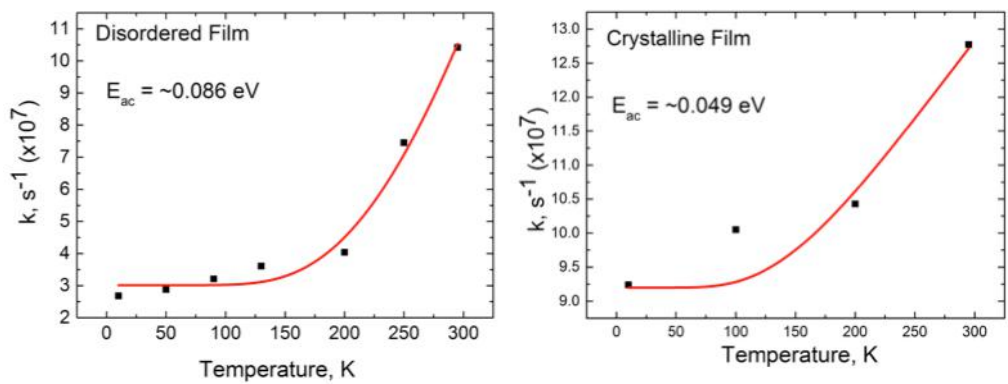

*Figure S8: The fluorescence rate ($s^{-1}$) at 650 nm plotted against temperature (K) for the TT emission in the disordered film (left) and polycrystalline film (right). We determine an activation energy of ~90 meV and ~50 meV respectively.*

**e.)Transient Grating Photoluminescence Spectroscopy**

Transient grating measurements were carried following a protocol set out in reference [2] on a set-up with a time resolution of 200 fs from 1 ps- 100 ps.

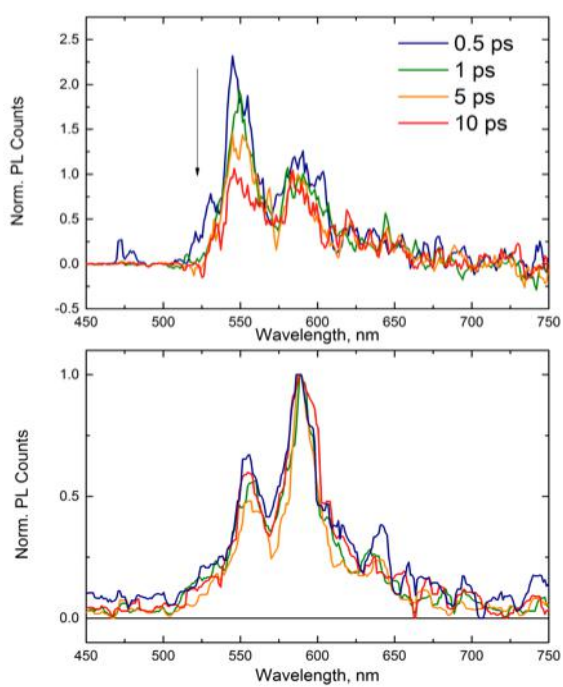

*Figure S9: Spectra taken at 0.5 ps- 10 ps, normalized to the maximum of the 0-1 peak, from the transient grating photoluminescence measurement of the disordered film (top) and polycrystalline film (bottom) at 295 K. We observe a loss in intensity in the 0-0 peak over the first 10 ps in the disordered film, but no change in the 0-0/0-1 peak ratio in the polycrystalline film.*

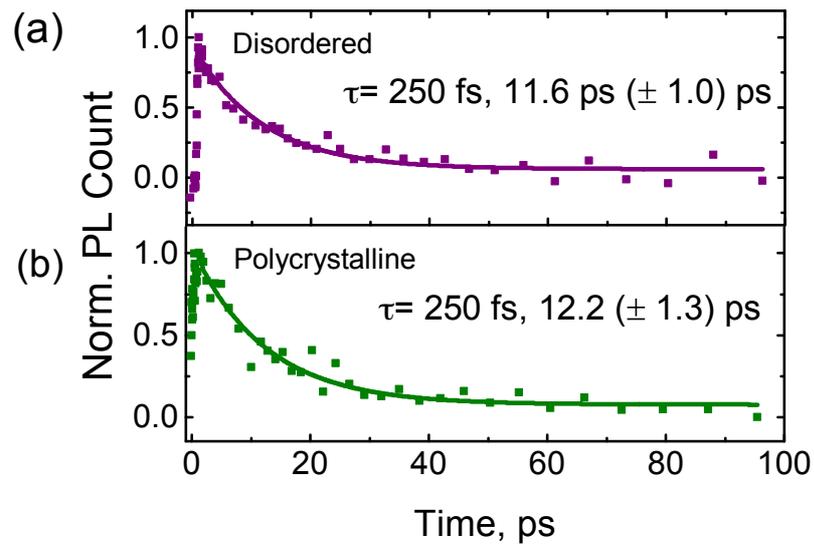

*Figure S10: Decay of $S_1$ PL at 540 nm for the disordered (a) and polycrystalline (b) films, measured on the TG set-up. Both decays are fit to a biexponential decay with $\tau_1$ set to 250 fs. We obtain $\tau_2$ fluorescence lifetimes of 11.6 ± 1 ps and 12.2 ±1.3 ps respectively.*

## 4.) Transient Absorption Spectroscopy

### a.) Fluence dependence

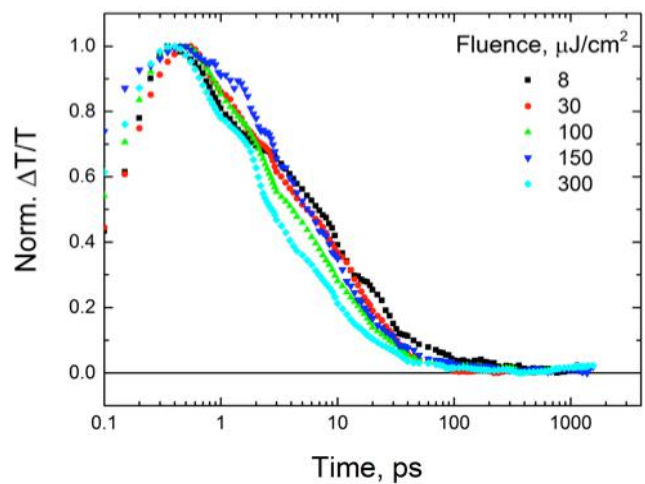

*Figure S11: $S_1$ decay in a disordered film, measured with a pump wavelength of 532 nm. We observe that the $S_1$ decay is unaffected by fluence up to ~ 100 uJ/cm$^2$ for both film types.*

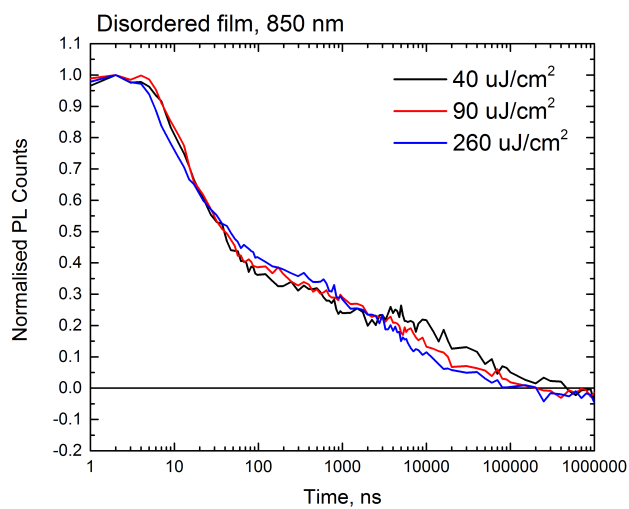

*Figure S12: Fluence-dependence of the transient absorption decay at 850 nm (TT and $T_1$ absorption peak) in the disordered film at 295 K, excitation wavelength= 532 nm. We observe little fluence-dependence in the disordered film before 100 ns, consistent with monomolecular decay. The decay of separated $T_1$ is accelerated slightly at higher fluence.*

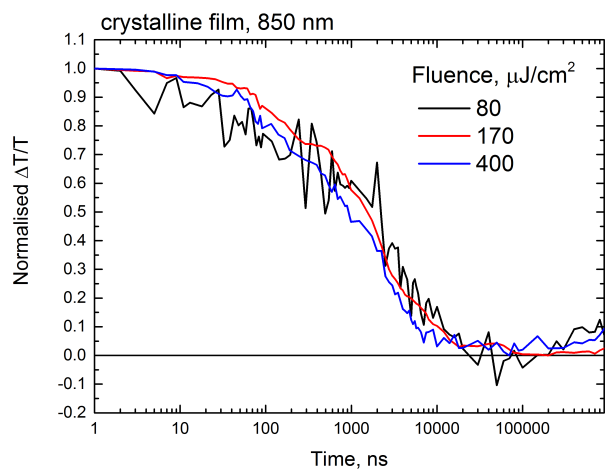

*Figure S13: Fluence-dependence of the transient absorption decay at 850 nm (TT and $T_1$ absorption peak) in the polycrystalline film at 295 K, excitation wavelength= 532 nm. The TT decay in the polycrystalline film shows no strong dependence on fluence up to 400 uJ/cm$^2$.*

## b.) Ultrafast TA spectra

TA spectra of TIPS-tetracene films were measured with 10 fs time resolution. Below we present the TA spectra, the residual maps recovered after globally fitting the TA maps and the FFT maps of the residuals.

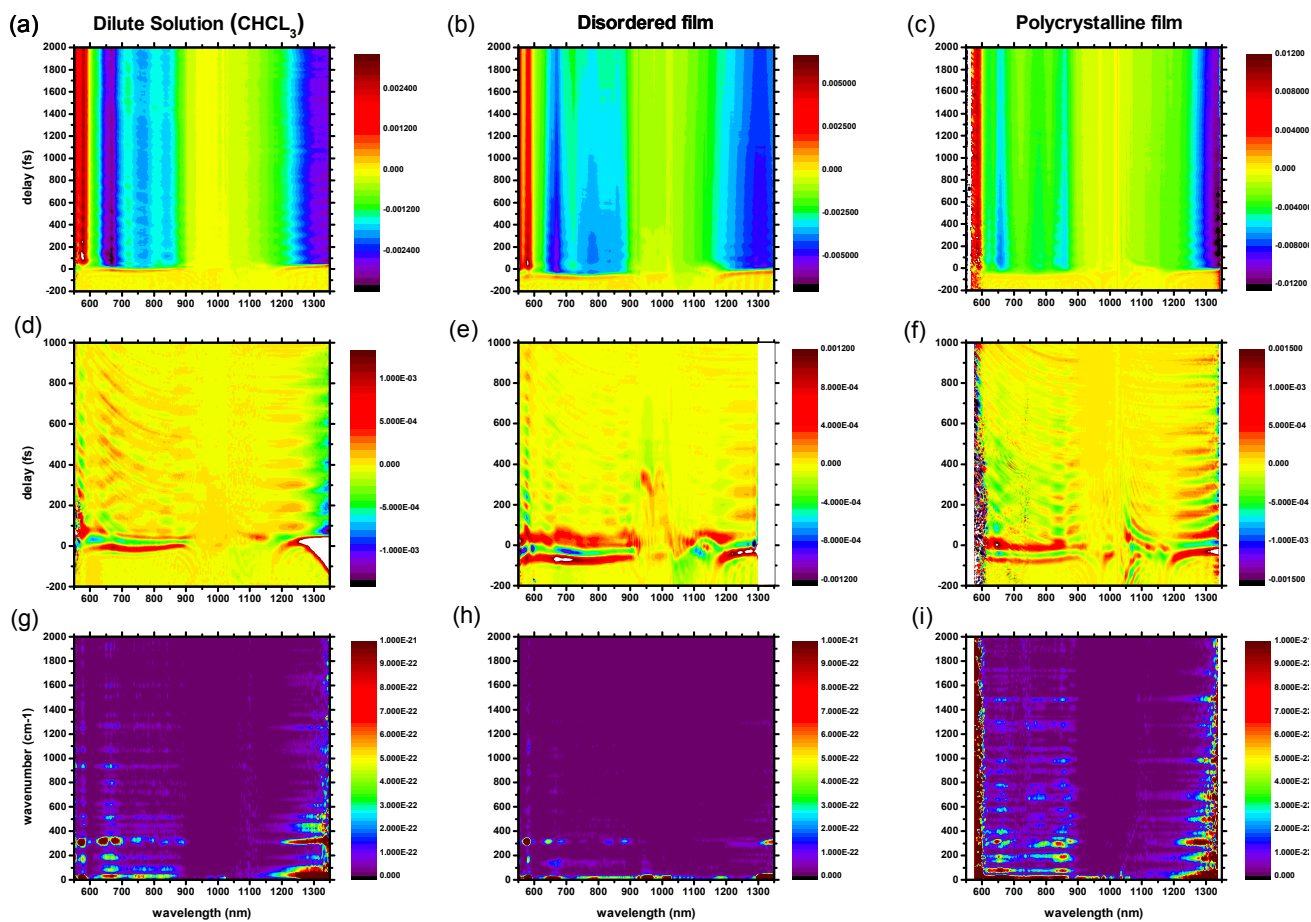

*Figure S14. (a,b,c) Ultrafast transient absorption spectra of dilute solution, a disordered film and a polycrystalline film. (d,e,f) The residuals maps of a-c recovered after globally fitting a-c and subtracting the electronic populations. (g,h,i) The FFT maps of measurements a-c.*

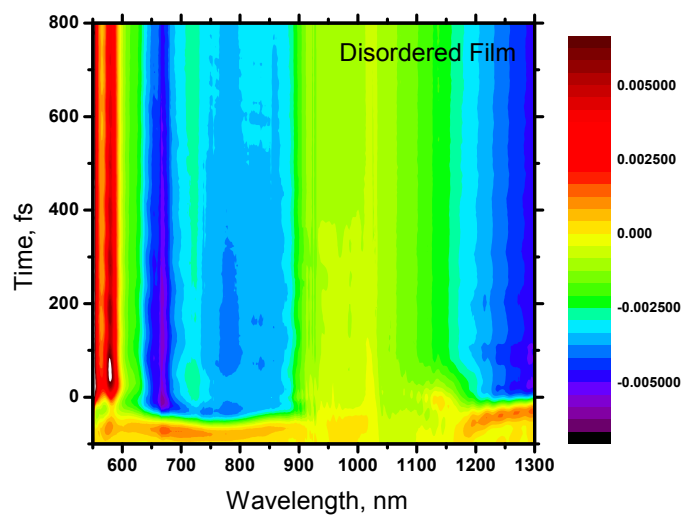

*Figure S15: Transient absorption measurement of the disordered film from -100 fs to 800 fs.*

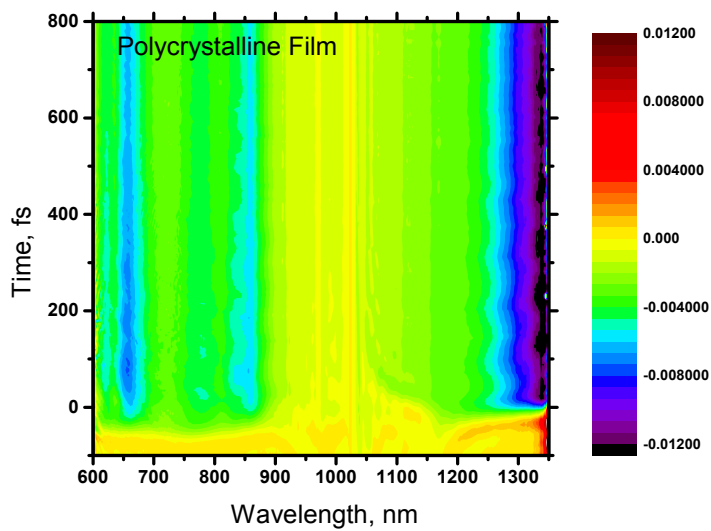

*Figure S16: Transient absorption measurement of the polycrystalline film from -100 fs to 800 fs.*

**c.) Ultrafast rise of the TT state (results obtained with spectral deconvolution algorithm).**

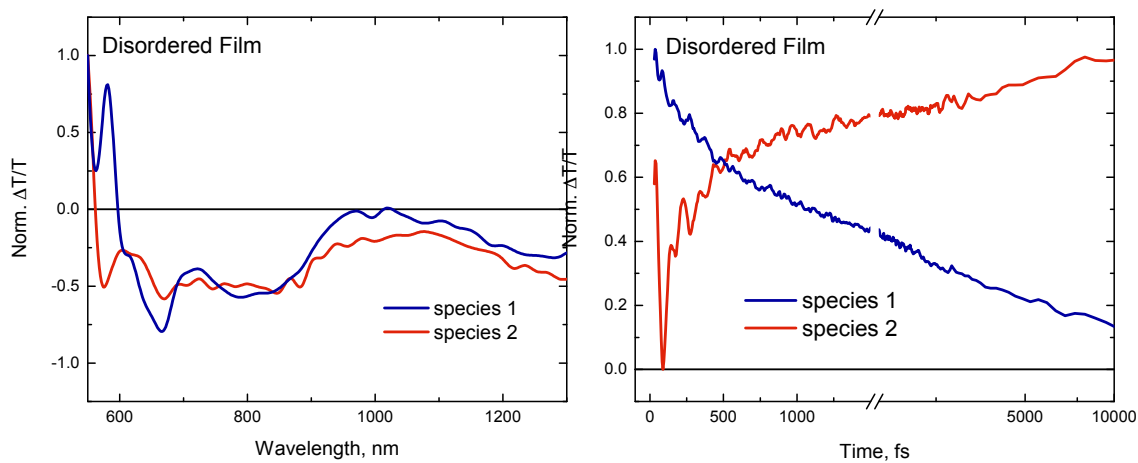

*Figure S17: The output of a spectra deconvolution algorithm run over the ultrafast TA measurement of the disordered film from 30 fs- 10 ps.*

### d.) Temperature-dependence

#### i.) S$_1$ ps decay

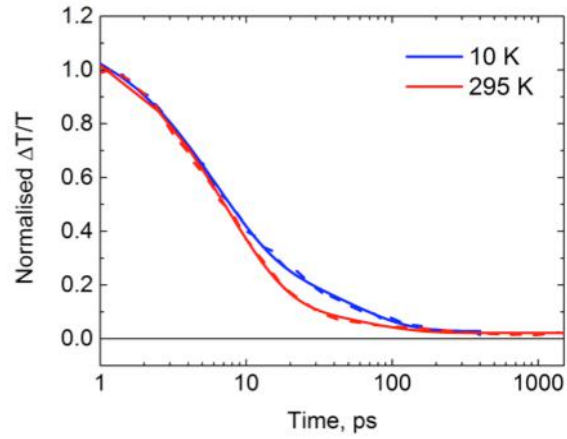

*Figure S18; Decay of S$_1$ in a disordered film from TA measurements at 10 K and 295 K. Lifetimes of 10 ps and 9 ps.*

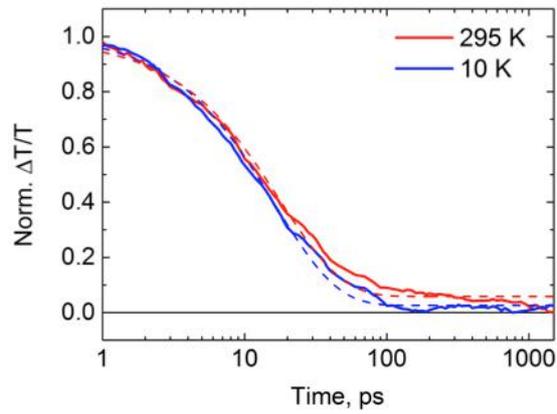

*Figure S19: Decay of S$_1$ in a polycrystalline film from TA measurements at 10 K and 295 K. Lifetimes of 16 ps and 17 ps.*

**ii.) TT dissociation**

We observe thermally-activated TT separation in the disordered film at 295 K. At 10 K we do not resolve any dissociation of the TT state. In the crystalline film we observe no dependence on temperature over the ns- ms timescale and resolve only the TT state absorption spectrum.

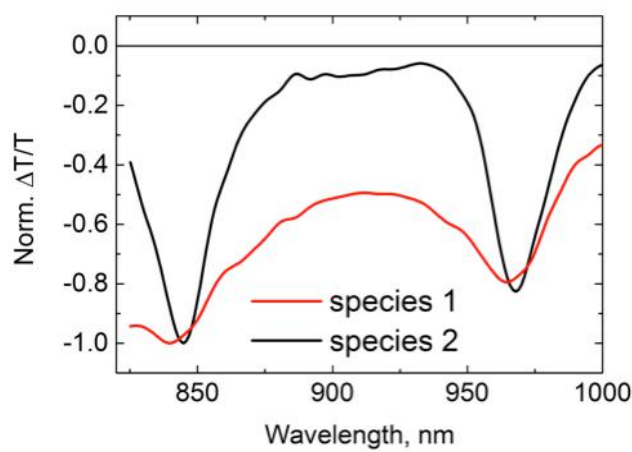

*Figure S20: The two spectral species identified in the 295 K ns-TA measurement of the disordered film. Species 1 shows a broadened absorption with the $T_1$ peaks blue-shifted by ~5 meV. Species 2 shows the $T_1$ absorption peaks observed in solution and confirmed to be $T_1$ via a separate sensitization technique.*

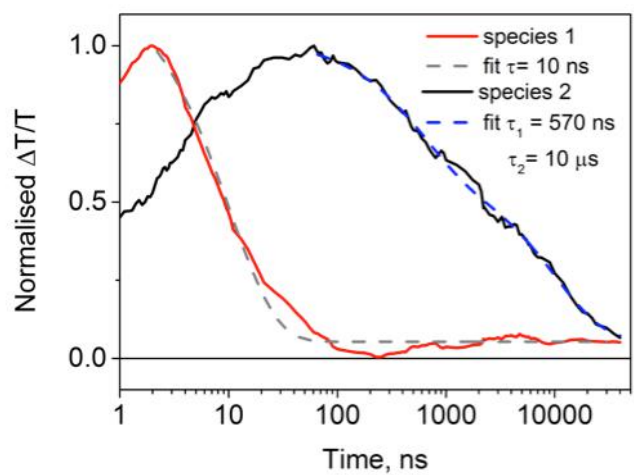

*Figure S21: The two kinetic components that correspond to the SVD spectral solutions for the 295 K ns-TA measurement of the disordered film. Species 1 represents the TT decay and shows a 10 ns lifetime. Species two corresponds to the decay of T1 and shows lifetimes of 570 ns and 10 μs.*

At 10 K we do not resolve TT separation in the disordered film; we identify only one spectral species using SVD (Fig S21), the TT state, and we observe no spectral shifting in the band position. The singular value spectra for the 10 K and 295 K nanosecond TA measurements, in the NIR spectral range, are shown below. There are two outlying components at 295 K, the TT and $T_1 + T_1$ and only one at 10 K.

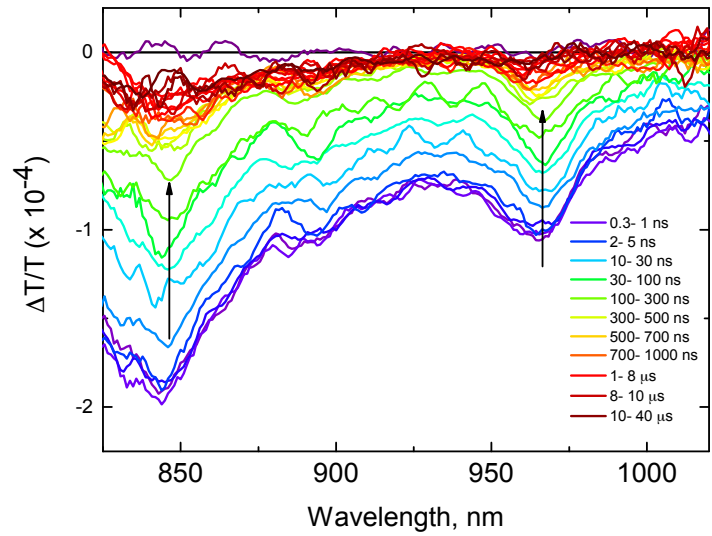

*Figure S22: The ns-TA spectra of a disordered film at 10 K.*

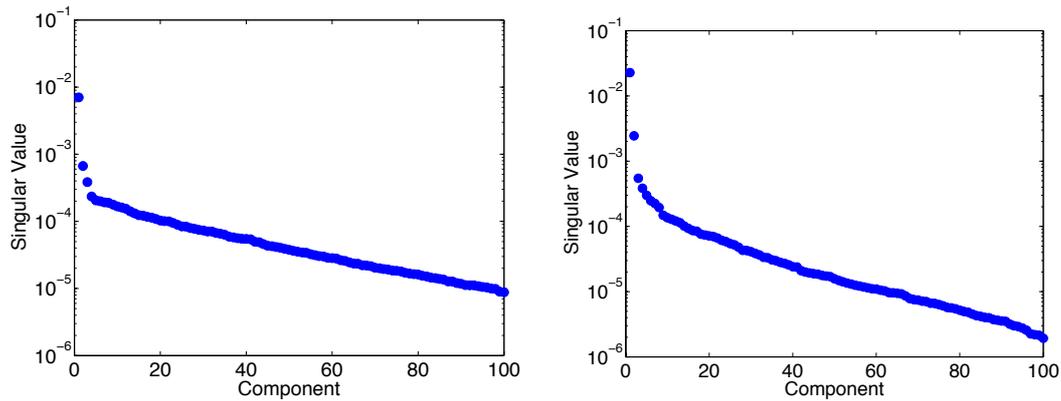

*Figure S23: The singular values from the SVD analysis of the disordered film at 10 K (left) and 295 K (right). The SVD only identifies one principal component at 10 K (the TT state) and two at 295 K (TT and $T_1 + T_1$).*

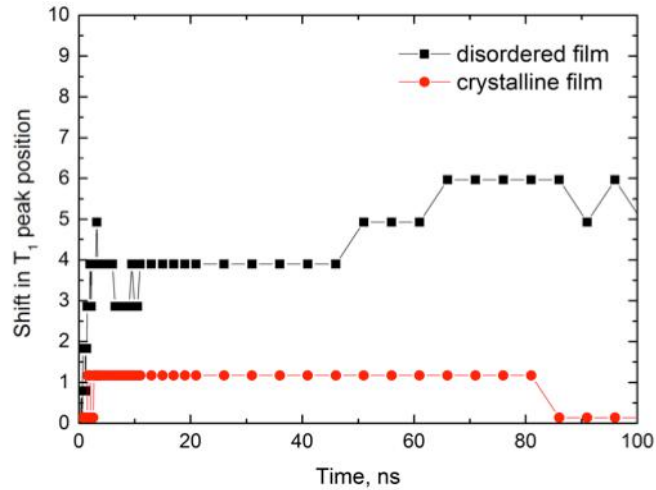

*Figure S24: The shift in $T_1$ peak position (966 nm peak) at 295 K in the disordered and polycrystalline films over the first 100 ns after excitation. The y axis is defined as the shift in peak position relative to the wavelength of the max intensity at 1 ns. We observe that the disordered film peak position shifts by ~ 6 nm (5 meV) over the first 10 ns as the TT dissociates into $T_1 + T_1$. In contrast, the max peak position in the crystalline film shifts one pixel over the first 100 ns. We note that the resolution of the TA measurement is ~1 nm.*

**e.) Spectral assignments**

*TT absorption*

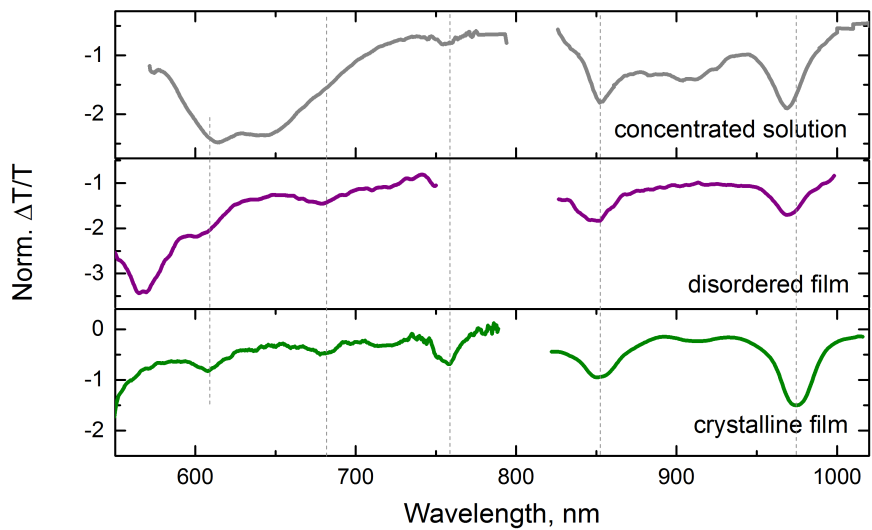

*Figure S25: The TT absorption spectrum identified by the spectral deconvolution algorithm for concentrated solution, the disordered film (295 K) and the polycrystalline film (295 K). The $T_1$ absorption bands displayed in the TT absorption are indicated with the dashed grey lines. We note the extra 900-920 nm TT absorption band is clear to see in the solution and crystalline material. It is broadened in the disordered film.*

*T₁ absorption*

The TIPS-tetracene $T_1$ absorption spectrum was confirmed in the solution study using a $T_1$ sensitization experiment with N-methylfulleropyrrolidine. For full details see reference [7]. In the solid-state we observe the same $T_1$ absorption bands as in solution.

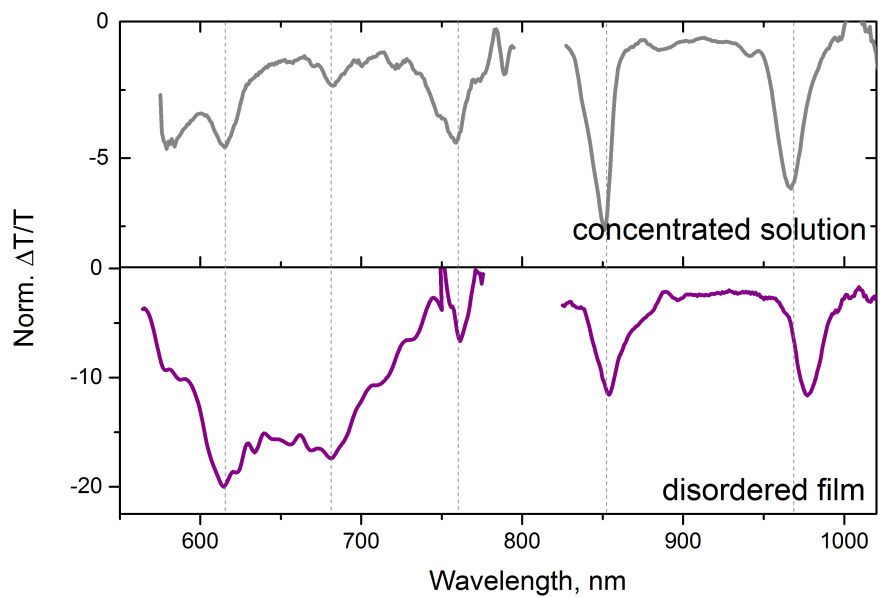

*Figure S26: The $T_1$ absorption spectrum identified by the spectral deconvolution algorithm for concentrated solution, and the disordered film. The $T_1$ absorption bands are indicated with the dashed grey lines.*

### 5.) Genetic algorithm for spectral deconvolution

To isolate the spectral species present in the transient absorption measurements of the TIPS-tetracene films, and to determine their individual evolution over time, we used a combination of singular value de-composition (SVD) and a spectral de-convolution code based on a genetic algorithm [8]. This code generates a given number of spectra that best reproduce the original data, while satisfying basic physical constraints such as spectral shape and population dynamics. The spectra are optimized by a genetic algorithm that minimizes the residual between the obtained spectra and the original data. In brief, the genetic algorithm is an example of an evolutionary algorithm that factors TA spectra into a pre-determined number of spectral components. The idea is based on natural selection of genes in a population; random spectra or parents are mixed until they give separate spectra that best reproduce the original data, which constitutes a test for genetic fitness. The output of the algorithm is a set of spectra and kinetics for the species identified.

We use this optimization method over other approaches because the algorithm requires minimal initial input; the only input is the number of species to optimize. The optimization starts from random initial spectra, if no initial solution is provided, and does not require a starting kinetic scheme, such as in global analysis methods. Unlike SVD, the algorithm is confined by physical constraints, such as non-negative kinetics. We use SVD to identify how many species are likely to be in the measurement.

The advantages of a genetic algorithm for our purpose over other optimization methods, such as gradient search methods, is that it is more capable of modeling a multidimensional data space that can be noisy and contain several overlapping features. Gradient search methods, on the other hand, are much more likely to get stuck in local minima. The schematic below demonstrates the operation of the algorithm:

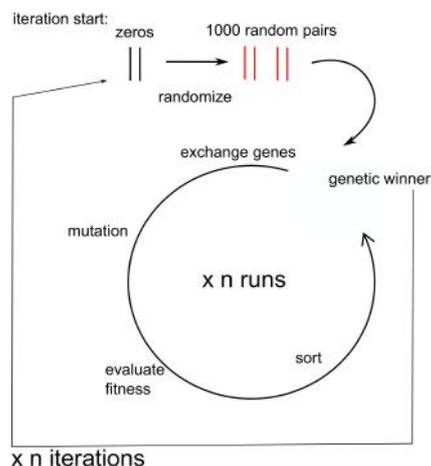

*Figure S27: Schematic demonstrating the general process of the genetic algorithm code.*

The algorithm starts with 1000 randomly generated set of vectors that are assessed for their fitness during one mutation cycle, which involves exchange of genes, mutation, assessment of fitness and sorting. We can specify the runs of times this mutation cycle occurs before the winning pair is taken out and re-randomized in a brand new iteration. This process prevents solutions being trapped in local minima. Once the algorithm has run for the specified number of iterations, the solutions are given.

### 6.) Computational Methods

### a.) Dimer Structures

The calculations presented were all performed on dimer or monomer structures. The geometry optimizations were performed using the ωB97X-D functional in the 6-31G* basis set. We chose the ωB97X-D functional because it includes Van der Waals interactions, which are the dominant type of intermolecular interactions present for the TIPS-tetracene molecule.

The dimer structure used for the disordered film assumes the optimum slip-stacked packing of two molecules, the same interaction that we assume forms in solution where molecules have diffusional freedom. In this case the geometry is optimized at the dimer level in the ground state.

The crystal structure was not optimized at a dimer level because the optimization would require a much larger set of molecules to properly model the structure. Instead, we optimized the monomer and placed it in maximum overlap with the crystal dimer structure. This was done for multiple unique dimers in the crystal, and we found that there are only two dimer structures with a significant intermolecular interaction. The results obtained from the two different crystal dimer geometries were essentially identical. The isolated dimer energies were modeled by separating the thin film geometry by 2 nm. The lowest $S_1$ and $T_1$ energies obtained from this geometry were identical to that of monomer calculations.

### b.) Triplet-Triplet Hopping

The $T_1$-$T_1$ hopping coupling is a two electron process and is therefore determined by Dexter energy transfer rates. The hopping rates depend on the overlap between the two different molecular wavefunctions, causing it to be a much more local effect than $S_1$ energy transfer.

To model the hopping rate we use the same dimer geometries. In the crystal, and more so the disordered film, this represents a maximum rate of energy transfer because any thermal fluctuations on the molecules and in their intermolecular separation will reduce the hopping rate.

We assume a Marcus hopping rate for the transfer and use reorganization energy of 0.33 eV [8]. The free energy change is assumed to be the TT binding energy in the thin film and crystal, 0.056 and 0.00 eV, respectively. This was chosen because we are most concerned with the initial separation of the TT state, and once one $T_1$ moves away there will be no appreciable binding in the TT state relative to two independent $T_1$. The couplings were obtained from linear response time-dependent density functional theory using the fragment excitation difference (FED) method with the B3LYP functional in the 6-31G* basis. This approach has been shown to give fairly accurate predictions of triplet diffusion lengths in organic crystals [8]. For the disordered film and crystal geometries we computed a $T_1$-$T_1$ coupling of 0.015 and 0.00017 eV, respectively. This gives a hopping time of 2.4 ps and 52 ns, respectively.

### 7.) Yield Approximation

The overall yield of $T_1 + T_1$ in the disordered film is temperature dependent. We estimate the yield at room temperature via two methods: using the photoluminescence (kinetic analysis) and from the relative strength of the $S_1$ and $T_1$ absorption. This second method uses the absorption extinction coefficients for the $S_1$ and $T_1$ on TIPS-tetracene that were determined in solution. Both methods require taking certain assumptions, thus they provide only an estimate of the yield.

### a.) Kinetic analysis

From analysis of the PLQE and emissive lifetime of the TT state at 10 K and 295 K, we can determine the radiative (k*rad*) and non-radiative (k*nrad*) decay rates. The PLQE value at 295 K (0.03 ± 0.005) was experimentally measured using an integrating sphere and the PLQE at 10 K (0.60± 0.1) was determined from the intensity increase of the steady-state emission from 295 K to 10K, which showed a x20 increase from 295 K to 10 K. We assume the difference in PLQE values with temperature reflects the change in radiative decay via the TT state, as the $S_1$ state has the same radiative lifetime at 10 K and 295 K and we observe rapid TT formation at both temperatures.

The rate of the decay of the TT state is given by the radiative decay rate (k*rad*), the non-radiative decay rate (k*nrad*) and the rate of TT dissociation to form $T_1 + T_1$ *(kdiss)*. Whilst the energy barrier from TT back to $S_1$ is lower than the barrier from TT to free triplets, reformation of $S_1$ will result in rapid singlet fission, reforming TT, while direct radiative or non-radiative decay of TT (e.g. free triplet generation) acts as a terminal loss pathway.

$$\frac{d(TT)}{dt} = -krad(TT)[TT] - knrad(TT)[TT] - kdiss(TT)[TT]$$

K*diss* is temperature dependent whilst we assume K*nrad* is not temperature dependent. This assumption is based on the observation in the crystalline material, where the TT state does not dissociate, is very weakly emissive at high and low temperature and shows the same lifetime at 10 K and 295 K. This suggests that the non-radiative decay pathways for the TT state in this material are relatively insensitive to temperature. For this calculation, we assume the same is true for the disordered material and give an upper bound for the dissociation rate.

At 10 K, there is no dissociation of TT into $T_1 + T_1$. Thus,

$$\frac{d(TT)}{dt} = -krad(TT)[TT] - knrad(TT)[TT]$$

We determine the krad and knrad decay rates from the PLQE and emission lifetimes. This is summarized in the following table:

|  | 10K | 295K |
|---|---|---|
|  |  |  |
| PLQE | $0.6 \pm 0.1$ | $0.03 \pm 0.005$ |
| $K_{rad}$ (s$^{-1}$) | $1.2\ (\pm 0.2) \times 10^7$ | $4.3\ (\pm 0.7) \times 10^6$ |
| $K_{nrad}$ (s$^{-1}$) | $7.9\ (\pm 1.5) \times 10^6$ | $7.9\ (\pm 1.5) \times 10^6$ |
| $K_{diss}$ (s$^{-1}$) | 0 | $1.3\ (\pm 0.2) \times 10^8$ |
| $K_{Total}$ (s$^{-1}$) | $2.0\ (\pm 0.5) \times 10^7$ | $1.4\ (\pm 0.2) \times 10^8$ |
| T (ns) | $50 \pm 10$ | $7 \pm 1$ |

We note that there is a slight difference in the calculated radiative rates for 10K and 295 K which is not expected.

We find that the K*diss* at 295 K represents ~90% of the total decay rate (error bars give 91-93% range), giving an estimate of the $T_1 + T_1$ yield of ~180-186%.

### b.) Sensitization

A second way we can estimate the $T_1 + T_1$ yield is using the absorption coefficients of the $S_1$ and $T_1$ states, determined by a solution sensitization experiment [7]. We assume that $S_1$ and $T_1$ display a similar absorption cross section in the solid-state as in solution. In some solid-state materials delocalization of the excitons make comparison with solution difficult. However, as pointed out in the main text, in TIPS-tetracene the similarity between solution and the solid-state for the absorption of $S_1$ and $T_1$ states suggests that excitons are relatively localized in the solid state. Therefore, we will use this method as a second estimate of the $T_1 + T_1$ yield.

$$\varepsilon[T_1] = -5463\ Lmol^{-1}cm^{-3}$$

And

$$\varepsilon[S_1] = 1.41 \times 10^4\ Lmol^{-1}cm^{-3}$$

If we consider a disordered film measurement in the main text (Figure 2(b)) and a 100 nm thick film:

$\frac{\Delta T}{T}(at\ 650\ nm, 1ps) = -0.002$

$[S^1] = 1.56 \times 10^{-2}\ molL^{-1}cm^{-1}$

$\frac{\Delta T}{T}(at\ 750\ nm, 500\ ns) = 1.0 \times 10^{-4}$
$[T^1] = 1.8 \times 10^{-2}\ molL^{-1}cm^{-1}$

$[T_1]/[S_1] = 130\%$

Clearly, the two yield estimation methods give a broad range (130-180%) for the $T_1 + T_1$ yield, reflecting the difficulty in determining this parameter. However, both show well over 100% $T_1$ generation. In addition, we note that the photoluminescence properties of TIPS-tetracene are not largely dissimilar from tetracene, where the $T_1$ yield is reported to be close to 200 %.